\newif\iffigs\figstrue
\figsfalse

\documentstyle[12pt]{article}
\iffigs
  \input epsf
\else
\fi

\batchmode
\newfont{\footscrfont}{rsfs10}
  \newfont{\footbbbfont}{msbm10}
  \newfont{\manfont}{manfnt}
\errorstopmode

\newif\ifscrf\scrftrue
\ifx\footscrfont\nullfont
  \scrffalse
\fi

\newif\ifamsf\amsftrue
\ifx\footbbbfont\nullfont
  \amsffalse
\fi

\def\ppnumber{\vbox{\baselineskip14pt\hbox{CU-TP-871}
\hbox{hep-th/9712054}}}
\def\ppdate{December 1997}
\def\pplogo{\vbox{\kern-\headheight\kern -15pt
\halign{##&##\hfil\cr&{%\sc
\ppnumber}\cr\rule{0pt}{2.5ex}&\ppdate\cr}
}}

\makeatletter
\date{}
\def\dedicatory#1{\def\@date{\normalsize\it#1}}
\def\subjclass#1{\def\@thefnmark{}\@footnotetext{1991
    {\it Mathematics Subject Classification.} #1}}
\def\keywords#1{\def\@thefnmark{}\@footnotetext{
    {\it Key words and phrases.} #1}}

\def\ps@firstpage{\ps@empty \def\@oddhead{\hss\pplogo}%
  \let\@evenhead\@oddhead % in case an article starts on a left-hand page
}
\def\maketitle{\par
 \begingroup
 \def\thefootnote{\fnsymbol{footnote}}
 \def\@makefnmark{\hbox
 to 0pt{$^{\@thefnmark}$\hss}}
 \if@twocolumn
 \twocolumn[\@maketitle]
 \else \newpage
 \global\@topnum\z@ \@maketitle \fi\thispagestyle{firstpage}\@thanks
 \endgroup
 \setcounter{footnote}{0}
 \let\maketitle\relax
 \let\@maketitle\relax
 \gdef\@thanks{}\gdef\@author{}\gdef\@title{}\let\thanks\relax}

\def\abstract{\if@twocolumn
\section*{Abstract}
\else \small
\begin{center}
{\bf ABSTRACT}
\end{center}
\quotation
\fi}

\def\thebibliography#1{\section*{References\@mkboth
 {REFERENCES}{REFERENCES}}\small\list
 {[\arabic{enumi}]}{\settowidth\labelwidth{[#1]}\leftmargin\labelwidth
 \advance\leftmargin\labelsep
 \usecounter{enumi}}
 \def\newblock{\hskip .11em plus .33em minus .07em}
 \sloppy\clubpenalty4000\widowpenalty4000
 \sfcode`\.=1000\relax}

\newif\iffn\fnfalse

\@ifundefined{reset@font}{\let\reset@font\empty}{} %needed for ancient LaTeX
\long\def\@footnotetext#1{\insert\footins{\reset@font\footnotesize
    \interlinepenalty\interfootnotelinepenalty
    \splittopskip\footnotesep
    \splitmaxdepth \dp\strutbox \floatingpenalty \@MM
    \hsize\columnwidth \@parboxrestore
   \edef\@currentlabel{\csname p@footnote\endcsname\@thefnmark}\@makefntext
    {\rule{\z@}{\footnotesep}\ignorespaces
      \fntrue#1\fnfalse\strut}}}

\makeatother

\ifamsf
  \newfont{\bigbbbfont}{msbm10 scaled\magstep2}
  \newfont{\bbbfont}{msbm10 scaled\magstep1}  % msbm12 does not exist
  \newfont{\smallbbbfont}{msbm8}
  \newfont{\tinybbbfont}{msbm6}
  \newfont{\smallfootbbbfont}{msbm7}
  \newfont{\tinyfootbbbfont}{msbm5}
  \newfont{\biggthfont}{eufm10 scaled\magstep2}
  \newfont{\gthfont}{eufm10 scaled\magstep1}  % eufm12 does not exist
  \newfont{\smallgthfont}{eufm8}
  \newfont{\tinygthfont}{eufm6}
  \newfont{\footgthfont}{eufm10}
  \newfont{\smallfootgthfont}{eufm7}
  \newfont{\tinyfootgthfont}{eufm5}
\fi

\ifscrf
  \newfont{\scrfont}{rsfs10 scaled\magstep1}  % rsfs12 does not exist
  \newfont{\smallscrfont}{rsfs7}
  \newfont{\tinyscrfont}{rsfs7}
  \newfont{\smallfootscrfont}{rsfs7}
  \newfont{\tinyfootscrfont}{rsfs7}
\fi

\ifamsf
  \newcommand{\Bbb}[1]{\iffn
      \mathchoice{\mbox{\footbbbfont #1}}{\mbox{\footbbbfont #1}}
      {\mbox{\smallfootbbbfont #1}}{\mbox{\tinyfootbbbfont #1}}\else
      \mathchoice{\mbox{\bbbfont #1}}{\mbox{\bbbfont #1}}
      {\mbox{\smallbbbfont #1}}{\mbox{\tinybbbfont #1}}\fi}
  
\else
  \def\bigbbbfont{\bf}
  \def\Bbb{\bf}
  
\fi

\ifscrf
  \newcommand{\Scr}[1]{\iffn
    \mathchoice{\mbox{\footscrfont #1}}{\mbox{\footscrfont #1}}
    {\mbox{\smallfootscrfont #1}}{\mbox{\tinyfootscrfont #1}}\else
    \mathchoice{\mbox{\scrfont #1}}{\mbox{\scrfont #1}}
    {\mbox{\smallscrfont #1}}{\mbox{\tinyscrfont #1}}\fi}
\else
  \def\Scr{\cal}
\fi

\def\operatorname#1{\mathop{\rm #1}\nolimits}
\def\C{{\Bbb C}}

\def\P{{\Bbb P}}

\def\Z{{\Bbb Z}}

\def\Pic{\operatorname{Pic}}

\def\Gr{\operatorname{Gr}}

\def\rank{\operatorname{rank}}
\def\bearray{\begin{eqnarray}}
\def\eearray{\end{eqnarray}}
\def\bearraynn{\begin{eqnarray*}}
\def\eearraynn{\end{eqnarray*}}
\def\bfig{\begin{figure}}
\def\efig{\end{figure}}

\def\opeq#1{\advance\lineskip#1 \advance\baselineskip#1
	\advance\lineskiplimit#1}

\def\cM{{\Scr M}}

\def\cD{{\Scr D}}

\def\cMc{{\hfuzz=100cm\hbox to 0pt{$\;\overline{\phantom{X}}$}\cM}}
\def\barcD{{\hfuzz=100cm\hbox to 0pt{$\;\overline{\phantom{X}}$}\cD}}

\ifamsf

\else

\fi

\def\Pf{{\em Proof: }}

\def\ker{{\rm ker}}
\def\coker{{\rm coker}}
\def\rank{{\rm rank}}
\def\im{{\rm im}}
\def\dim{{\rm dim}}
\def\codim{{\rm codim}}
\def\Card{{\rm Card}}

\def\deg{{\rm deg}}
\def\det{{\rm det}}
\def\Div{{\rm Div}}
\def\supp{{\rm supp}}
\def\Gr{{\rm Gr}}

\newtheorem{Proposition}{Proposition}[section]
\newtheorem{Definition}{Definition}[section]
\newtheorem{Theorem}{Theorem}[section]
\newtheorem{Lemma}{Lemma}[section]
\newtheorem{Corrolary}{Corrolary}[section]

\newcommand{\be}{\begin{equation}}
\newcommand{\ee}{\end{equation}}
\newcommand{\bp}{\begin{Proposition}}
\newcommand{\ep}{\end{Proposition}}
\newcommand{\bt}{\begin{Theorem}}
\newcommand{\et}{\end{Theorem}}
\newcommand{\bl}{\begin{Lemma}}
\newcommand{\el}{\end{Lemma}}
\newcommand{\bc}{\begin{Corrolary}}
\newcommand{\ec}{\end{Corrolary}}

\begin{document} 

\title{On degree zero semistable bundles over an elliptic curve}

\author{C.~I.~Lazaroiu$^a$}
\maketitle
\vbox{
\centerline{Department of Physics} 
\centerline{Columbia University}
\centerline{New York, NY 10027}
\medskip
\bigskip
}

\abstract{Motivated by the study of heterotic string compactifications 
on elliptically fibered Calabi-Yau manifolds,
we present a procedure for testing semistability and identifying 
the decomposition type of degree zero holomorphic vector bundles over a 
nonsingular elliptic curve. The algorithm requires explicit knowledge 
of a basis of sections of an associated `twisted bundle'.}

\vskip 3.5 in

$^a$ lazaroiu@phys.columbia.edu

\pagebreak

\section{Introduction}

In the study of a certain class of heterotic string compactifications one 
encounters the following: 

{\bf Problem:}
{\em Given a smooth elliptic curve $E$ and a degree zero holomorphic 
vector bundle $V$ over $E$, find a practical algorithm to 
determine whether $V$ is semistable.
In this case, find a maximal decomposition of $V$ in indecomposable 
subbundles. }

This question appeared in \cite{CGL} in the course of an investigation of the 
relation between $(0,2)$ heterotic string compactifications and $F$-theory.

Since the most accessible bundle data are usually 
its global holomorphic sections, we will be interested in solving this problem
by finding a characterization of semistability and of the decomposition type 
in terms of properties of a basis of sections of a certain bundle 
associated to $V$. Our main result is Theorem \ref{final_theorem} in section
2.3.

{\bf Physical motivation:}
A $(0,2)$ heterotic compactification is characterized by a Calabi-Yau manifold
$Z$ and a stable holomorphic vector bundle $V$ over $Z$ \cite{Witten_old}.
If one is interested in models having a potential $F$-theory dual 
\cite{V,VM,FMW}, one takes
$Z$ to be elliptically fibered and with a section. In this case, for
a certain component of the moduli space, there exists an alternate 
description of stable vector bundles over $Z$ in terms of pairs
$(\Sigma,L)$ where $\Sigma$ is the spectral cover of $V$ and $L$ is a line 
bundle over $\Sigma$ \cite{FMW,FMW_math,Donagi_Markman}.
Such data are easier to manipulate then the
abstract bundle data. On the other hand, there exists an accessible class of 
$(0,2)$ 
compactifications, namely those realized via $(0,2)$ linear sigma models 
\cite{Witten_phases,Distler_notes,MDG,CDG}.
In this case, $V$ is presented as the sheaf cohomology of a monad defined
over $Z$, while $Z$ itself is realized as a complete intersection in a 
toric variety \cite{BB}. This leads to the problem, studied in \cite{CGL}, 
of translating between these alternative 
presentations of $V$ in the $(0,2)$ linear case.
The main condition for $V$ to admit a spectral cover 
description is that its restriction $V|_E$ to the generic elliptic fibre $E$ 
of $Z$ be semistable. In order to carry out the task of \cite{CGL}, one needed
a method to test this condition for a given bundle $V$. This proves
essential in organizing the wealth of models that can be built. An important
point, which was tangentially mentioned in \cite{CGL}, is that the 
above condition often fails to hold, even for $(0,2)$ models which seem to be 
physically well-defined. One also finds a significant number of models 
for which the condition is satisfied but $V|_E$ does not fully decompose as a 
direct sum of line bundles. Discriminating between such cases can be achieved
by the methods of the present paper. On the other hand, the method of 
\cite{CGL} was justified only for the case when $V|_E$ is semistable and 
fully decomposable. Here we remedy this by providing a systematic discussion 
of the general situation. 

{\bf Mathematical context:}
The main results we need date back to a classical paper of Atiyah 
\cite{Atiyah}
\footnote{Background material can be found for example in \cite{Potier}.} 
.
Fix a nonsingular elliptic curve $E$ with a distinguished point $p$.
Let ${\cal E}(r,0)$ be the set of (holomorphic equivalence classes of)
{\em indecomposable} holomorphic vector bundles of rank $r$ and degree zero 
over $E$. Any element $V \in {\cal E}(r,0)$ is of
the form $V=L\otimes F_r$ with $L\in \Pic^0(E)$ a degree zero line bundle 
uniquely determined by $V$ and satisfying $L^r=\det V$. 
Here $F_r$ is the unique
element of ${\cal E}(r,0)$ with $h^0 \neq 0$. One has $h^0(F_r)=1$.
The bundles $F_r$ can be defined inductively by $F_1:=O_E$  and by the
fact that $F_r$ is the unique nontrivial extension :
\be
0\longrightarrow F_{r-1}\longrightarrow F_r\longrightarrow
O_E\longrightarrow 0
\ee
of $F_{r-1}$ by $O_E$. The Riemann-Roch theorem gives
$h^1(F_r)=h^0(F_r)=1$. It is known that $F_r$ is semistable for all
$r$.

For any holomorphic vector bundle of degree zero and rank $r$ over
$E$, consider a maximal decomposition as a direct sum of holomorphic
subbundles:
\be
V=\oplus_{j=1..k}{V_j}
\ee

If $V$ is semistable, then we necessarily have $\deg V_j\le 0$ for all
$j=1..k$ and since $0=\deg V=\sum_{j=1..k}{\deg V_j}$, it follows that
$\deg V_j=0$ for all $j$. If $r_j:=\rank V_j$, we thus have 
$V_j \in {\cal E}(r_j,0)$ and $V_j = L_j \otimes F_{r_j}$, with 
$L_j \in \Pic^0(E)$. Thus :
\be
\label{splitting}
V=\oplus_{j=1..k}{L_j \otimes F_{r_j}}
\ee
Note that $ \sum_{j=1..k}{r_j}=r$.

Conversely, if such a decomposition of $V$ exists, then, since all terms are 
semistable and of slope $0$, a standard result 
(see \cite[p17,Cor.~7]{Seshadri}) assures us that $V$ is 
semistable and of degree zero. The idea of our approach will be to use 
(\ref{splitting}) in order to simultaneously check semistability and 
determine the maximal decomposition, thus avoiding the difficult problem of 
testing semistability independently.

The sequence of pairs $(r_j,L_j) (j=1..k)$ will be called the 
{\em decomposition type} (or {\em splitting type}) of $V$. 
By using the distinguished point $p \in E$ to write $L_j \approx O(q_j-p)$ 
for some
$q_j \in E$, we can identify this data with the sequence of pairs 
$(r_j,q_j)$, modulo the choice of $p$. Obviously the splitting type 
determines $V$ up to isomorphism. 

Part of this information  is encoded by what we will call the 
{\em spectral divisor} $\Sigma_V$ of $V$, defined by :
\be
\Sigma_V:=r_1q_1+...+r_kq_k \in \Div(E)
\ee
Note that some of the points $q_j$ may coincide. If $q_1=...=q_{j_1}:=Q_1~$,
 ... ,$~q_{j_1+...+j_{l-1}+1}=..=q_{j_1+...+j_l}:=Q_l$ 
(with $j_1+...+j_l=k$), then 
$\Sigma_V=\rho_1Q_1+...+\rho_l Q_l$ where 
\be
\nonumber
\rho_i=\sum_{j_1+...+j_{i-1}+1 \le i \le j_1+...+j_i}{r_i}.
\ee
In particular, $\Sigma_V$ cannot discriminate between direct 
factors of the type 
$O(Q_1)\otimes(F_{r_1}\oplus ... \oplus F_{r_{j_1}})$ and factors 
of the type $O(Q_1)\otimes F_{r_1+...+r_{j_1}}$. In fact, it is easy to see
\footnote{Since the only stable bundles of slope zero over an elliptic 
curve are the degree zero line bundles, any Jordan-Holder (JH) filtration of 
$V$ is by subbundles of consecutive dimension. The (isomorphism class of) the
associated graded bundle  
$gr(V)$ is independent of the choice of the JH filtration. 
If $V$ decomposes as above, the natural JH filtrations of $F_{r_i}$ 
induce a JH filtration of $V$ in the obvious way. The associated graded 
bundle is $gr(V)=O(Q_1-p)^{\oplus \rho_1}\oplus...\oplus 
O(Q_l-p)^{\oplus \rho_l}$. Therefore, $\Sigma$ depends only on $gr(V)$, i.e. 
only on the $S$-equivalence class of $V$.}
that $\Sigma_V$ depends only on the $S$-equivalence class of 
$V$. Two degree zero semistable vector bundles having the same spectral 
divisor need not have the same splitting type.

The explicit computation of $\Sigma_V$ was the main task of \cite{CGL}.
In that paper, a solution of this problem was presented only for the 
`fully split' case (this is rigorously formulated in Section 3). A by-product 
of the study we undertake here is a simple generalization of the method of 
\cite{CGL} for determining the spectral divisor 
(see Corrolary \ref{spectral_div} in section 2.3).

We will often consider the `twisted' bundle $V':=V \otimes O(p)$, 
which has degree $r$ and slope $1$.
If $V$ is semistable, one has the following 

\begin{Lemma}
\label{cohom_dim}
Let $V$ be a degree zero semistable vector bundle over $E$. 
Then $h^0(V')=\rank V$ and $h^1(V')=0$.
\end{Lemma}

{\em Proof:}
By (\ref{splitting}), we have $h^0(V')=
\sum_{j=1..k}{h^0(O(q_j) \otimes F_{r_j})}$.
Since $O(q_j) \otimes F_{r_j}$ is indecomposable and of positive degree, 
a result 
of \cite{Atiyah} shows that $h^0(O(q_j) \otimes F_{r_j})=
\deg O(q_j)\otimes F_{r_j}=r_j$ and the Riemann-Roch theorem gives 
$h^1(O(q_j) \otimes F_{r_j})=0$. This implies the conclusion.
$\Box$

As input data for the resolution of our problem we will assume explicit 
knowledge of a basis of sections of $V'$. This is typically easily computed,
at least if $V$ is presented as the sheaf cohomology of a monad. \footnote{
Indeed, in that case one can consider the $O(p)$-twisted monad. 
The long exact 
cohomology sequence of the twisted monad will collapse due to the fact that 
$h^1(V')=0$. This is one of the nice properties of $V'$.}

The plan of this paper is as follows. 
In section 2 we study semistability and the decomposition type for a degree 
zero
holomorphic vector bundle $V$ over $E$. We formulate necessary and sufficient 
conditions on a basis of sections of $V'$ in order for $V$ to be 
semistable; this will also indicate its decomposition type. 
In particular, we obtain a simple receipt for the spectral 
divisor. We also consider the spectral divisor in the monad case and 
propose a `moduli problem'. 

In section 3 we consider the fully decomposable (`fully split') case. 
We present a criterion for
identifying fully decomposable and semistable vector bundles of degree zero 
over $E$, together with an algorithmic implementation. This is the main case 
considered in \cite{CGL}. The novelty here is that the algorithm we 
give tests semistability of $V$ (and at the same time determines its 
spectral divisor and its decomposition type, thus describing $V$ completely 
in the language of \cite{Atiyah}); 
in \cite{CGL}, the focus was on computing $\Sigma_V$ and 
$V$ was {\em assumed} to be semistable and fully decomposable 
in order to simplify the presentation.  
We also explain how one can analyze $V$ by starting from more 
general twists. This is necessary in practice in cases when one
cannot easily compute the sections of bundles over $E$ twisted by  $O(p)$.
\footnote{In the set-up of \cite{CGL}, one is interested in smooth 
elliptic curves 
realized as complete intersections in a toric variety $\P$. 
In this case, one can easily compute the sections of $V\otimes L_E$, for 
restrictions $L_E$ of 
reflexive sheaves $L$ over $\P$. 
If $O(p)$ is not such a restriction then the 
sections of $V \otimes O(p)$ are not easily accessible. 
For example, if $E$ is realized as a cubic in $\P^2$, the line bundle
$O_{\P^2}(1)$ over $\P^2$ restricts to a degree three line bundle 
$O(p_1+p_2+p_3)$ over $E$,  
and one can apply the methods of section 3 to the twisted bundle 
$V':=V\otimes O(p_1+p_2+p_3)$.}.
With the physics oriented reader in mind, the discussion of section 3 is 
carried out by a direct approach and can be 
read independently of the rest of the paper; it is intended as a technical 
companion of \cite{CGL}.

{\bf Notation and terminology:} If $s$ is a regular section of a holomorphic 
vector bundle, 
then $(s)$ denotes the zero divisor (divisor of zeroes) of $s$. 
$\Div(E)$ is the free abelian 
group of divisors on $E$. If $D \in \Div(E), D=\sum_{j=1..k}{n_j p_j}$, with 
$n_j \in \Z, p_j \in E$, then $\supp D$ denotes the set $\{p_j|j=1..k\}$. 
All vector bundles and their morphisms are holomorphic. For any vector space
$A$, $\Gr ^k(A)$ denotes the grassmanian of $k$-dimensional subspaces of $A$.
If $S \in A$ is a subset, then $<S>$ denotes the linear span of $S$.  
For any holomorphic bundle $R$ over $E$, $\Gr ^k(R)$ denotes the set of 
rank $k$ holomorphic subbundles of $R$. $\sim$ denotes linear equivalence of 
divisors and $\Pic(E)$ the Picard group of $E$. If $r$ is an integer, then 
$\Pic^r(E)$ is the set of isomorphism classes of degree $r$ line bundles over 
$E$; it is only a sub{\em set} of $\Pic(E)$, except for $r=0$, when it is a 
subgroup. We say that a filtration $0=K_0\subset K_1\subset ... \subset K_r=U$
of an $r$-dimenisonal vector space $U$ is {\em nondegenerate} if 
$K_{i-1} \neq K_i$ for all $i=1..r$. Then $K_i$ have consecutive dimensions. 
For a holomorphic bundle $V$, 
we denote by $\mu(V):=\deg V/\rank V$ its slope (nomalized degree).

{\bf Intuitive idea}
The starting point for our analysis is the fact that the 
twisted bundles $F'_r$ are given recursively as nontrivial extensions 
of $O(p)$ by itself. By Lemma \ref{cohom_dim}, the associated cohomology 
sequences collapse and this provides a very good handle on the behaviour 
of $F'_r$. In terms of the local behaviour of sections, the difference between 
$F'_r$ and the completely trivial extension $O(p)^{\oplus r}$ is manifest only 
at the point $p$. In both cases, the bundles admit a basis of $r$ sections 
whose values are linearly independent at each point of $E$ except $p$. At 
this point, the behaviour in the two cases is dramatically different. While 
in the completely decomposable case the values of all sections vanish 
simultaneously at $p$ along linearly independent `directions', in the case 
of $F'_r$ only one of them vanishes, while the others have linearly 
independent values. In the latter case, however, the `direction' of the first 
section approaches the space spanned by the values of the others as we 
approach $p$ on $E$, and at the point $p$ it lies in that space. 
The behaviour of the 
sections of $V'$ can be obtained essentially by a `linear superposition' from 
the behaviour of its indecomposable factors. 
Most of what follows 
consists in developing enough technology in order to make these ideas precise.
This being understood, the physics-oriented reader may at first consider 
only the 
first part of subsection 2.1, the statements of Theorems \ref{F_criterion} and 
\ref{final_theorem} in section 2 and of Theorem \ref{theorem_fully_split} 
in section 3 and the associated algorithm. 
 
\section{General analysis}

	Let $V$ be a degree zero holomorphic vector bundle over a smooth 
elliptic curve $E$. Fix a point $p \in E$ and define $V':=V\otimes O(p)$.

	We present a criterion for deciding whether $V$ is
semistable and, in this case, for determining its splitting type.
This criterion requires explicit knowledge of a basis of holomorphic 
sections of $V'$.

	The plan of this section is as follows. In subsection 1 we discuss
a notion of order of incidence of a holomorphic section on a subbundle.  
Since this discussion does not require assuming $\deg V=0$, we will
present it for a general holomorphic vector bundle over $E$. In subsection 2
 we use these concepts to describe the sections of the bundles
$F'_r$. In subsection 3 we give our characterization of
degree zero semistable bundles.

\subsection{Incidence order of holomorphic sections on subbundles} 

	In this subsection let $W$ be a rank $r$ holomorphic vector bundle 
over $E$ and let $T$ be a rank $r_0$ holomorphic subbundle of $W$.

	Any {\em nonzero} regular section of $W$ defines a unique line 
subbundle
$L_s$ of $W$ in the following way (see \cite{Atiyah}).
For each $t \in \supp (s)$, choose
a local holomorphic coordinate $z$ on $E$ centered at $t$. 
Let $\nu_t$ be the degree of vanishing of $s$ at $t$. Then 
$\exists \lim_{e\rightarrow p}{z^{-\nu_t}s(e)} :={\hat s}(t)$, 
where ${\hat s}(t) \in W_t - \{ 0 \}$.
We define $(L_s)_e:=<s(e)>$, for all $e \in E - \supp (s)$ and 
$(L_s)_t:=<{\hat s}(t)>$ for all $t \in \supp (s)$. Note that changing the local 
holomorphic coordinate $z$ to another local holomorphic coordinate $z'$ 
centered at $t$ will change 
${\hat s}(t)$ to  ${\hat s}'(t)=\lim_{e \rightarrow p}
(z(e)/z'(e))^{\nu_t}{\hat s}(t)$.
Thus, the vector  ${\hat s}(t)$ is defined up to multiplication by a 
nonzero complex number. In particular, $(L_s)_t$ is 
well-defined. By using the local triviality of $W$ or by the argument given in 
\cite{Atiyah}, one can convince oneself that $L_s$ is a 
holomorphic subbundle of $W$.
Note that $L_{\lambda s}=L_s$, $\forall \lambda \in C^*$, so that
we have a well-defined map $\P H^0(W) \longrightarrow \Gr ^1(W)$ from
the projectivisation of $H^0(W)$ to the set of holomorphic line
subbundles of $W$.

	Since $s$ is a holomorphic section of $L_s$, it follows that $L_s$ is 
holomorphically equivalent to $O(s)$,
where $O(s)$ is the line bundle on $E$ associated to the divisor 
$(s) = \sum_{t \in \supp (s)}{\nu_t t}$. In particular, we have 
$\deg L_s =\deg (s) =\sum_{ t \in \supp (s)}{ \nu_t }=
\sum_{e \in E}{\deg s(e)}$, where we define $\deg s(e)$ to be $\nu_e$, if 
$e \in \supp (s)$ and $0$ otherwise.

	For each $e \in E$, we have a natural linear map 
$\phi_e :H^0(W) \longrightarrow W_e$ given by 
$\phi_e(s):=s(e),\forall s \in H^0(W)$ (the evaluation map at $e$). 
We denote its image and kernel by 
$R_e:=\phi_e(H^0(W)) \subset W_e$, 
$K_e:=\ker \phi_e \subset H^0(W)$ 
and we  define 
$r_e(W):=\dim_{\C} R_e$, 
$d_e(W):=\dim_{\C} K_e$.
We have $r_e(W)+d_e(W)=h^0(W)$ at any point $e \in E$. 

Define a subspace $N_e$ of $W_e$ by 
$ N_e:=<\{{\hat s}(e)|s \in K_e\}> \subset W_e$ 
(if $s=0$, we define ${\hat s}(e)$ to be zero). 
It is easy to see that changing the 
linear coordinate $z$ does not affect $N_e$. Note that 
$N_e=\sum_{s\in K_e}{(L_s)_e}$. 
In general, the subspaces $N_e,R_e$ 
of  $W_e$ may intersect and their sum need not generate $W_e$.
Define ${\cal Z}(W):=\{t \in E | K_t \neq 0\}$. 

	If $W$ is semistable then we must have $\deg s=\deg L_s \le \mu(W)$. 
Since $s$ is regular, we also have $\deg s \ge 0$ . 
Then $\deg s \in \{ 0,..,[\mu(W)] \}$, where $[~]$ denotes the integer part. 
In particular, we have $\deg s(e) \le \mu(W)$ for all $e \in E$.

\begin{Proposition}
\label{isom}
Suppose that $W$ is semistable and of slope $1$ .
Let $e \in E$ and fix a local coordinate $z$  around $e$ on $E$. 
Then the map $s \in K_e \rightarrow {\hat s} \in N_e$ is a $\C$-linear 
isomorphism. In particular, we have $\dim_{\C} N_e=d_e$.
\end{Proposition}

{\em Proof:} By the above, we see that any $s \in K_e -\{0\}$ must have 
a {\em simple} zero at $e$. If $s_1,s_2 \in K_e$ and 
$\alpha_1,\alpha_2 \in \C$, let $s:=\alpha_1 s_1 + \alpha_2 s_2$. Then 
$\exists \lim_{e' \rightarrow e}{z^{-1}s(e')}=\alpha_1 {\hat s}_1(e) + 
\alpha_2 {\hat s}_2(e)$. If 
$\alpha_1 {\hat s}_1(e) + \alpha_2 {\hat s}_2(e)=0$,
then $s$ must be zero (otherwise $s$ would have degree $>1$ at $e$). 
In this case ${\hat s}(e)=0$ by definition. If 
$\alpha_1 {\hat s}_1(e) + \alpha_2 {\hat s}_2(e)\neq 0$, then 
${\hat s}(e)=\alpha_1 {\hat s}_1(e) + \alpha_2 {\hat s}_2(e)$.

Thus in both cases we have 
${\hat s}(e)=\alpha_1 {\hat s}_1(e) + \alpha_2 {\hat s}_2(e)$,
which shows linearity. If $s \in K_e$, then by definition 
${\hat s}(e)$ is zero only if $s=0$. This shows injectivity. 
Surjectivity is obvious. $\Box$

What follows is a generalization of the previous classical discussion.

\begin{Definition}

	Let $s \in H^0(W)$ be a holomorphic section of $W$.
Consider the holomorphic section ${\overline s}$ of the 
quotient bundle $W/T$, naturally induced by $s$. We say that $s$ is 
{ \em incident of order (degree) $d$ on $T$ at a point $e \in E$} if 
${\overline s}$ has a 
zero of order (exactly) $d$ at $e$. In this case, we write $\deg _Ts(e):=d$ 
and we call it {\em the incidence order(degree)} of $s$ on $T$ at e. 

\end{Definition}

	Note that we have $s(e) \in T_e$ iff $\deg _Ts(e) > 0 $. Intuitively,
$\deg _Ts(e)$ characterizes `how fast' $s(e') \in W_{e'}$ approaches the 
subspace $T_{e'}$ of $W_{e'}$ as $e'$ approaches $e$ on $E$.  
	
	If $s \in H^0(T) \subset H^0(W)$, then ${\overline s}$ is identically 
zero, so the degree of incidence of $s$ on $T$ is not defined for such $s$ 
at any point of $E$. If $s \in H^0(W) - H^0(T)$, then ${\overline s}$ is a 
nonzero section of $W/T$ and the associated divisor $({\overline s})$ 
is a finite set of points of $E$.
Therefore, the set $Z_T(s):=\{ e \in E | \deg _Ts(e) >0 \}=
\{ e \in E | s(e) \in T \}=\supp ({\overline s})$ is finite for all sections 
$s\in H^0(W) - H^0(T)$. In particular,
$\deg _Ts(e)$ is well defined in this case at all points $e \in E$. Thus, for 
all $s \in H^0(W) - H^0(T)$, we can define the {\em total degree of $s$ 
along T} by  $\deg _Ts:=\sum_{e \in Z_s(T)}{\deg _Ts(e)}=\deg {\overline s}$.

	For $T={\bf 0}$ (the null subbundle of $W$) we have 
${\overline s}=s$ so 
$\deg _{\bf 0}s(e)=\deg s(e) $ and the above definition reduces
to the usual one.

\begin{Proposition}
\label{induced_section}
Let $M$ be a holomorphic subbundle of $T$ and $s \in H^0(W) -H^0(T)$.
Let $q \in E$ an arbitrary point. Let $\sigma$ be the section of $W/M$ 
induced by $s$ via the canonical projection $W\stackrel{p}{\rightarrow}W/M$.
Then $\deg _{T}s(q)=\deg _{T/M}\sigma(q)$
\end{Proposition}

{\em Proof:} Oviously $T/M$ is a subbundle of $W/M$ and $\sigma \in 
H^0(W/M)-H^0(T/M)$. $s$ and $\sigma$ induce the same section ${\overline s}$
of $W/T$ via the canonical projections $W\rightarrow W/T$ and 
$W/M \rightarrow (W/M)/(T/M)\approx W/T$.
Therefore:
$\deg _{T/M}\sigma(q)=\deg {\overline s}(q)=\deg _Ts(q)$. 
$\Box$

	We have the following :

\begin{Proposition}
\label{degree_bound}
	Suppose that $W$ is semistable of normalized degree $\mu(W)$ and that
$T$ has normalized degree $\mu(T)=\mu(W)$. 
Then we have $\deg _Ts \leq \mu(W)$ for all $s \in H^0(W) - H^0(T)$. 

\end{Proposition}

{\em Proof:}	Indeed, $T$ is in this case obviously semistable 
(since $W$ is semistable) and thus $W/T$ is semistable of normalized degree 
$\mu(W/T)=\mu(W)$ (see, for example Proposition 8 on page 18 of 
\cite{Seshadri}).
Then ${\overline s}$ must have have total degree at most equal to 
$\mu(W/T)=\mu(W)$ in order for  
$L_{\overline s}\approx O(s)$ not to destabilize 
$W/T$. Then use $\deg _Ts=\deg {\overline s}$. $\Box$

	For $W$ semistable and $\mu(T)=\mu(W)=1$, this 
shows that a section $s \in H^0(W) - H^0(T)$ either does not 
intersect $T$ or intersects it at exactly one point, the incidence 
degree of $s$ at that point being exactly one.

	We now give an alternative description of the incidence degree, 
which is more practical from a computational point of view.

\begin{Proposition}

	Let $s \in H^0(W) - H^0(T)$ and $q \in E$.
Consider a local holomorphic frame $(s_1 ... s_{r_0})$ of $T$ around $q$.
Then 
\be
\label{frame_criterion}
\deg s(q)=\deg s\wedge s_1\wedge ... \wedge s_{r_0}(q) 
\ee

\end{Proposition}

{\em Proof:}
Let $U$ be an open neighborhood of $q$ such that the exact sequence

\be
\label{split_sequence}
0 \longrightarrow T|_U \stackrel {j}{\longrightarrow} W|_U 
\stackrel{p}{\longrightarrow}(W/T)|_U 
\longrightarrow 0
\ee
splits in the holomorphic category. Let $u :(W/T)|_u \longrightarrow W|_U$ be 
a holomorphic injection such that 
$W|_U=j(T|_U) \oplus u((W/T)|_U)$. 
We identify $T$ with $j(T)$ via $j$ and $(W/T)|_U$ with 
$u((W/T)|_U)$ 
via $u$.  We can assume that $U$ is small enough so that all 3 bundles
involved are trivial above $U$.
Let $s_1...s_{r_0}$ be a local holomorphic frame of $T$ above 
$U$ and $s_{r_0+1}...s_r$ a frame of 
$(W/T)|_U\equiv u((W/T)|_U)$. Then $s_1...s_r$ is a local frame 
of $W$ above $U$. 

	Write $s(e)=\sum_{i=1..r}{f_i(e)s_i(e)}$ with $f_i \in {\cal O}_U$. 
Then 
\be
\nonumber
{\overline s}(e)=\sum_{i=r_0+1...r}{f_i(e)s_i(e)}. 
\ee
and
\be
\nonumber
s(e)\wedge s_1(e) \wedge ... \wedge s_{r_0}(e)=
\sum_{i=r_0+1...r}{f_i(e)s_i(e) \wedge s_1(e) \wedge ... \wedge s_{r_0}(e)} 
\ee

The statement 
$\deg _Ts(q)=d$ is equivalent to 
$\exists \lim_{e \rightarrow q}{z^{-d}{\overline s}(e)} \neq 0$, 
which is equivalent to 
$\exists \lim_{e \rightarrow q}{z^{-d}{\overline f}(e)} \neq 0$, 
where 
${\overline f}:= (f_{r_0+1}...f_r) \in \oplus_{i=r_0+1 ...r}{{\cal O}_U}$.
This in turn is equivalent to $\exists \lim_{e \rightarrow q}{z^{-d}s(e)
\wedge s_1(e) \wedge ... \wedge s_{r_0}(e)}\neq 0$.
$\Box$

	Now let $s \in H^0(W) -H^0(T)$ and $q \in E$.
The associated section ${\overline s} \in H^0(W/T)$ defines a line 
subbundle 
$L_{\overline s} \subset W/T$ as above. 
In particular, at 
the point $q$ we have a $1$-dimensional subspace $(L_{\overline s})_q$ of 
the fibre $(W/T)_q=W_q/T_q$.
We define $W_s(q)$ to be the $(r_0+1)$-dimensional subspace of $W_q$ 
which 
induces $(L_{\overline s})_q$, i.e. the
preimage of $(L_{\overline s})_q$ via the natural surjection 
$W_q\stackrel{p_q}{\longrightarrow} W_q/T_q $. For $e \in E -Z_T(s)$ 
we obviously have $W_s(q)=<s(e)>\oplus T_e$. The following gives 
an analogue of this decomposition for points $ q \in Z_T(s)$:

\begin{Proposition}
\label{osc_vector}
	Let $s \in H^0(W) - H^0(T) $ and $q \in E$. 
Let $z$ be any local holomorphic coordinate on $E$, centered at $q$.

	The following are equivalent :

(a) $\deg _Ts(q)=d$

(b) There exist {\em local} holomorphic sections ${\tilde s}$ of $W$ 
around $q$
and ${s_0}$ of $T$ around $q$ such that :
	
	(b1) $s(e)=z^d{\tilde s}(e) + s_0(e) $ for all $e$ sufficiently close 
	     to $q$

	(b2) ${\tilde s}(q) \in W_q - T_q $

In this case, we have $W_s(q)=<{\tilde s}(q)> \oplus T_q$.

\end{Proposition}

{\em Proof:} 
Assume $(a)$ holds and 
consider a neighborhood $U$ of $q$ such that the 
sequence \ref{split_sequence} splits. Since $\deg {\overline s}(q)=d$, 
we can choose $U$ small enough so that there exists  
a holomorphic section $\sigma$ of $W/T$ above $U$ such that 
${\overline s}(e)=z^{d}\sigma(e), \forall e \in U$ and
$\sigma(q) \neq 0$.  
Then there exists a holomorphic section ${\tilde s}:=u \circ 
\sigma$ of $W|_U$, 
such that ${\overline {\tilde s}}=p({\tilde s})=\sigma$. Thus  
$p(s-z^d{\tilde s})=0$, so that $s(e) -z^d {\tilde s}(e) \in T_e$, 
$\forall e \in U$. Since $s(e) -z^d {\tilde s}(e)$ is holomorphic, 
this gives a holomorphic section $s_0$ of $T|_U$ such that 
$s=z^d{\tilde s}+s_0$ and $(b1)$ holds. 
Moreover, $\sigma(q) \neq 0$ implies 
${\tilde s}(q) \in W_q -T_q$ and thus
$(b2)$ holds.
Since ${\overline s}(e)=z^d \sigma(e)$, we have 
${\hat {\overline s}}(q)=\sigma(q)$ so that 
$\sigma(q) \in (L_{\overline s})_q$. Thus 
${\tilde s}(q) \in p_q^{-1}((L_{\overline s})_q)=W_s(q)$ and 
$W_s(q)=<{\tilde s}(q)> \oplus T_q$.

The converse implication is trivial in view of the previous proposition. 
$\Box$

Note that ${\tilde s}$, $s_0$ cannot, 
in general, be extended beyond a neighborhood of $q$. 
Also note that ${\tilde s}(q)$ is only determined modulo $T_q$ and 
modulo a constant multiplicative factor 
(from the choice of the local holomorphic coordinate $z$ around $q$).

\begin{Definition}

Let $s \in H^0(W)-H^0(T)$. Define 
$W_{s,T}=\sqcup_{e \in E}{W_s(e)}$. 
Then $W_{s,T}$ has a natural structure of holomorphic vector bundle over $E$
and $s \in H^0(W_{s,T})$ while $T$ is a holomorphic subbundle of $W_{s,T}$.

\end{Definition}
{\em Proof:}
A holomorphic trivialization of $W_{s,T}$ is obtained as follows. For $U$ 
an open set such that $U \cap Z_T(s)=\Phi$, choose a local frame 
$s_1..s_{r_0}$ of $T$ over $U$ and trivialize $W_{s,T}$ over $U$ by using the 
local frame $s_1...s_{r_0},s$. For $U$ such that $U \cap Z_T(s)={q}$ 
( a single point), by choosing $U$ small enough and picking a local 
holomorphic coordinate $z$ on $E$, one one can write $s(e)=z^d{\tilde s}(e) + 
s_0(e)$ as before, where $d=\deg _T(s)$ and ${\tilde s}$ does not meet $T$ 
over $U$. 
Then ${\tilde s}$ is a local 
holomorphic
section of $W$ above $U$ and one can trivialize $W_{s,T}$ over 
$U$ by using
$s_1...s_{r_0},{\tilde s}$, where $s_1...s_{r_0}$ is a local holomorphic frame
of $T$ above $U$. The holomorphic 
compatibility of the various local trivializations is immediate.
$\Box$

Intuitively, the fibre $W_{s,T}(q)=<{\tilde s}(q)> \oplus T_q$ for 
$q \in Z_T(s)$ is the correct `limit' of the fibres 
$W_{s,T}(e)=<s(e)> \oplus T_e$ as $e \rightarrow q$. The section $s$ 
determines a line subbundle $L_s$ of $W$ and we have $W_{s,T}=L_s\oplus T$. 
For $T={\bf 0}$
(the null subbundle of $W$), we obviously have $W_{s,{\bf 0}}=L_s$. 
This is a generalization of the construction of $L_s$.

Now suppose that the set ${\cal Z}(W)$ is finite
\footnote{We will see that this is the case if $W=V\otimes O(p)$ with 
$V$ semistable and of degree zero}. 
In this case, if $s_1...s_k \in H^0(W)$ are $\C$-linearly independent 
sections of 
$W$, then
they are also $\C$-linearly independent at the generic point of 
$E$ (i.e. $s_1(e)...s_k(e)$ are linearly independent in $W_e$ for
a generic $e \in E$); then we can define
inductively $W_{s_1...s_k}:=W_{s_k,W_{s_1...s_{k-1}}}$, with 
$W_{s_1}:=L_{s_1}$. Indeed, one can easily see that 
$s_2 \in H^0(W) -H^0(W_{s_1})$ and (by induction) 
$s_j \in H^0(W)-H^0(W_{s_1...s_{j-1}})$, 
$\forall j=2..k$, due to the generic 
linear independence of $s_1..s_k$. $W_{s_1...s_k}$ is a rank $k$ 
vector bundle and $s_1..s_k$ are sections of $W_{s_1..s_k}$ which are 
linearly independent at the generic point. Intuitivley, $W_{s_1...s_k}$ is 
the subbundle of $W$ `spanned' by $s_1..s_k$. 

If $(\sigma_1...\sigma_k)^t=A(s_1...s_k)^t$, with $A \in GL(k,\C)$ a constant
nondegenerate matrix, then it is easy to see that $W_{\sigma_1..\sigma_k}=
W_{s_1...s_k}$. Indeed, $s_1 ... s_k$ are sections of 
$W_{\sigma_1..\sigma_k}$ (since $\sigma_1...\sigma_k$ are) so that 
$W_{\sigma_1...\sigma_k,e}=W_{s_1..s_k,e}$ for all $e$ with $s_1(e)...s_k(e)$
linearly independent. For $q\in E$ such that $s_1(q)...s_k(q)$ are linearly 
dependent, one can consider vectors 
${\tilde \sigma}_1(q),..,{\tilde \sigma_k}(q)$, 
determined by local sections $\sigma_j$ of $W_{\sigma_1...\sigma_j}$ and 
$\sigma_{0j}$ of  $W_{\sigma_1...\sigma_{j-1}}$ via the conditions:    
$\sigma_j(e)=z^{deg_{W_{\sigma_1...\sigma_{j-1}}}\sigma_j(q)}
{\tilde \sigma}_j(e)+
\sigma_{0j}(e)$ for $e$ close to $q$ and 
$\sigma_j(q) \in W_{\sigma_1..\sigma_{j},q}-W_{\sigma1...\sigma_{j-1},q}$.
Note that ${\tilde \sigma}_1(q),..,{\tilde \sigma_k}(q)$ are linearly 
independent. 
These vectors obviously belong to $W_{s_1...s_k}(q)$ since for $e \neq q$ 
they are related to 
$\sigma_1...\sigma_k$ (and thus to $s_1...s_k$) by linear 
combinations of these vectors and since subbundles of $W$ are closed 
in the total space of $W$. 

Thus $W_{\sigma_1...\sigma_k}(q)=
<{\tilde \sigma}_1(q),..,{\tilde \sigma_k}(q)> \subset W_{s_1..s_k}(q)$ 
and they must coincide since they have the same dimension. Therefore, 
$W_{s_1...s_k}$ depends only on the subspace $<s_1...s_k>$ of $H^0(W)$.
Thus, {\rm if} ${\cal Z}(W)$ {\rm is finite} then we have a natural map :

\be
\psi_k:\Gr ^k(H^0(W))\rightarrow \Gr ^k(W).
\ee

An alternative way to understand this is as follows (cf. \cite{Atiyah}). 
If ${\cal Z}(W)$ is 
finite, then given a $k$-dimensional 
subspace $K$ of 
$H^0(W)$, $\phi_e(K)$ defines a rational section $f$
of  ${\Bbb G}{\rm r}^k(W)$, where ${\Bbb G}{\rm r}^k(W)$ is 
the bundle obtained by taking the grassmannian $\Gr^k(W_e)$ of $W_e$ as 
the fibre above each $e \in E$. Singularities of this section may appear 
only at a point $e$ where $\phi_e(K)$ fails to be $k$-dimensonal, i.e. 
at the points $e_1..e_s$ of $E$ where the values of a system $s_1..s_k$ 
of sections of $W$ giving a 
basis of $K$ fail to be linearly independent. Loosely speaking, 
one may worry that at such points 
there is no `completion' of the set $\{\phi_e(K)| e \in E-\{e_1..e_s\}\}$
which makes it into the total space of a holomorphic vector bundle. 
This does not happen for the following reason. 
With the natural structure, 
${\Bbb G}{\rm r}^k(W)$ is a complete variety and a classical result  
implies that $f$ must be regular. Thus $f$ determines a 
subbundle of $W$, which clearly coincides with $W_{s_1...s_k}$.

Again assuming ${\cal Z}(W)$ to be finite, 
suppose that we are given a filtration 
${\cal K}: 
0:=K_0 \subset K_1 \subset ...\subset K_{k-1} \subset K_k$ 
of a subspace $K_k$ of  $H^0(W)$, such that 
$\dim _{\C}K_j=j$, $\forall j=0..r$.
Associated to ${\cal K}$ via $\psi$ there is a filtration 
${\cal W}({\cal K}):
0:=W_0 \subset W_1 \subset ...\subset W_{k-1} \subset 
W_k$ 
by holomorphic subbundles with $\rank W_j=j$, $\forall j=0..r$.
If $s_j \in K_j-K_{j-1}$ for all $j=1..k$, then it 
is obvious that the integers 
$\delta^{\cal K}_j(t):=
\deg s_1\wedge...\wedge s_j(t)$ $(t \in {\cal Z}(W), j=1..r)$
depend only on ${\cal K}$. 
It is also easy to see -- by using Proposition 
\ref{osc_vector} -- that 
$\deg _{W_{j-1}}s_j(t) = 
\delta^{\cal K}_j(t)-\delta^{\cal K}_{j-1}(t)$, where we let 
$\delta^{\cal K}_0(t)$ be equal to $0$.

\subsection{The space of sections of the bundles $F'_r$}
	
	Let $F'_r:=F_r \otimes O(p)$. $F'_r$ is a rank $r$ indecomposable 
and semistable bundle of slope $1$. Since $F_r$ is semistable and of degree
zero, we have $h^0(F'_r)=r$ and 
$h^1(F'_r)=0$. Recall from \cite{Atiyah} that we have exact sequences:

\be
0\longrightarrow F_{k} \stackrel{i}{\longrightarrow}F_r 
\stackrel{p}{\longrightarrow} F_l \longrightarrow 0
\ee
for all $k,l \geq 0$ with $k+l=r$.
Below we will use their twisted version :
\be
\label{twisted}
0\longrightarrow F'_{k} \stackrel{i}{\longrightarrow}F'_r 
\stackrel{p}{\longrightarrow} F'_l \longrightarrow 0
\ee
For $l=1$, we obtain the twisted version of the defining sequences of $F'_r$: 
\be
\label{twisted_1}
0\longrightarrow F'_{r-1} \stackrel{i}{\longrightarrow}F'_r 
\stackrel{p}{\longrightarrow} O(p) \longrightarrow 0
\ee
while for  $k=1$ this gives :
\be
\label{twisted_2}
0\longrightarrow O(p) \stackrel{j}{\longrightarrow} F'_r \stackrel{p}
{\longrightarrow} F'_{r-1}\longrightarrow 0 
\ee

Since $H^1(F'_k)=0$, 
the exact cohomology sequence associated to (\ref{twisted}) 
collapses to:
\be
\label{short}
0\longrightarrow H^0(F'_{k}) \stackrel{i_*}{\longrightarrow}H^0(F'_r) 
\stackrel{p_*}{\longrightarrow} H^0(F'_l) \longrightarrow 0
\ee
Being an exact sequence of vector spaces, this must split. Therefore, there 
must exist $\C$-bases $<\sigma_1...\sigma_k>$ of $H^0(F'_{k})$, 
$<\sigma_{k+1}...\sigma_r>$ of $H^0(F'_l)$ and $<s_1...s_r>$ of $H^0(F'_r)$
such that $j_*(\sigma_i)=s_i, \forall i=1..k$ and 
$p_*(s_j)=\sigma_j, \forall j=k+1..r$.

\begin{Proposition}
\label{ker_prop}
For any $r\geq 1$, we have $d_p(F'_r)=1$ and 
$d_e(F'_r)=0$ for all $e \in E-\{p\}$.
\end{Proposition}

{\em Proof:}
The sequence (\ref{twisted_2}) shows that $d_p(F'_r) > 0$.

Now suppose that $d_p(F'_r)>1$. Then there exist two linearly independent
sections $s_1,s_2$ of $F'_r$ such that $s_1(p)=s_2(p)=0$. Let $L_i=L_{s_i}$ 
be the associated line subbundles of $F'_r$. Since $F'_r$ is semistable and 
of degree 1, we must have $\deg L_i=1$ and $(s_i)=p_i$. 
Hence $\exists \lim_{e \rightarrow q}{s_i(e)/z}=
{\hat s}_i(q) \neq 0$. 

Suppose that ${\hat s}_1(p),{\hat s}_2(p)$ are linearly dependent. 
Then we can write ${\hat s}_1(p)=\alpha {\hat s}_2(p)$ with $\alpha \in 
\C^*$. The section $s:=s_1 -\alpha s_2$ is then nonzero (since $s_1,s_2$ are 
$\C$-linearly independent) and we obviously have $\deg s(p)\ge 2$, 
which contradicts semistability
of $F'_r$. Thus, it must be the case that ${\hat s}_1(q),{\hat s}_2(q)$ are 
linearly independent. 

Now suppose there exists $e_0 \in E-\{p\}$ such that $s_1(e_0)$ and 
$s_2(e_0)$ are
linearly dependent. Write $s_1(e_0)=\beta s_2(e_0)$, with $\beta \in \C^*$. 
Then the section $s'=s_1 - \beta s_2$ vanishes both at $e_0$ and at $p$ 
and so $\deg s'(p)\ge 2$, again contradicting semistability of $F'_r$. 
It follows 
that $s_1(e), s_2(e)$ are linearly independent for all $ e \in E -\{p\}$.

From these two facts we immediately see that the subbundle sum $L_1 + L_2$ 
is {\em direct}. Since $(s_i)=p$, we also have $L_i\approx O(p)$; thus 
we have a holomorphic subbundle $L_1 \oplus L_2 =O(p) \oplus O(p)$ of 
$F'_r$.
Twisting by $O(-p)$, this gives a trivial subbundle of rank two 
$I_2 \subset F_r$. Since $h^0(I_2)=2$, this would imply $h^0(F_r)\geq 2$, 
a contradiction. This finishes the proof of the first statement. 

Now let $e \in E-\{p\}$. To show that $K_e=0$, we 
proceed by induction on $r$, using the sequence (\ref{twisted_2}).

For $r=1$ the statement is obvious. Suppose the statement holds for $r-1$, but
fails for $r$. Then there exists a nonzero section $s$ of 
$F'_r$ such that $s(e)=0$. We cannot have 
$s \in H^0(\im j)$ since that would imply 
$\deg s \geq 2$ (as $e \neq p$), 
which contradicts semistability of 
$F'_r$. Thus ${\overline s}: = p_*(s)$ is a nonzero section of $F'_{r-1}$.
Since $s(e)=0$, we have ${\overline s}(e)=0$, so that $K_e(F'_{r-1}) \neq 0$. 
This is impossible by the induction hypothesis. 
$\Box$

Consider the commutative 
group structure $(E,\oplus)$ on $E$ with zero element $p$. 
If $q_1,q_2 \in E$, then  $q_1\oplus q_2$ is defined to be the unique 
point $q$ of $E$ such that $(q_1)+(q_2)\sim (q)+(p)$, i.e. 
$O(q_1+q_2)\approx O(q+p)$. Thus $(q_1\oplus q_2)\sim (q_1)+(q_2)-(p)$.
Then $(q_1\oplus ...\oplus q_r) \sim (q_1)+...+(q_r)-(r-1)p$.

Let $T_r^{(p)}(E)$ be the $r$-torsion
subgroup of $(E,p)$, i.e. the set of points $t \in E$ such that 
$rt=0$ in $(E,\oplus)$, which is equivalent to $r(t)\sim r(p)$, i.e 
$O(rt)\approx O(rp)$. The map $q \in E \rightarrow O(q-p) \in \Pic^0(E)$
is a group isomorphism from $(E,\oplus)$ to $\Pic^0(E)$, which maps 
$T_r^{(p)}(E)$ to the subgroup 
$U_r:=\{L \in \Pic^0(E)| L^r\approx O_E\}\subset \Pic^0(E)$ of 
roots of order $r$ of $O_E$. We have $U_r \approx (\Z_r)^2$.

\begin{Proposition}

Let  $r >0$. The isomorphism classes of {\em indecomposable} 
bundles $A'$ which can be presented 
as extensions : 
\be
0\longrightarrow I_{r-1} \stackrel{j}{\longrightarrow} A' 
\stackrel{p}{\longrightarrow} O(rp)\longrightarrow 0
\ee
of
$O(rp)$ by the trivial rank $r-1$ bundle $I_{r-1}$ are in bijective  
correspondence with $U_r$. More precisely, each such bundle $A'$ is of the 
form:
\be
A'=O(q)\otimes F_r=L\otimes F'_r
\ee
where $q \in T_r^{(p)}(E)$ and $L:=O(q-p)\in U_r$. Here 
$F'_r:=F\otimes O(p)$.

\end{Proposition}

Note that we are {\em not} considering extension classes, but isomorphism 
classes of bundles which can be presented as extensions.

{\em Proof:} 

{\em Show that $F'_r$ fit into such sequences}

Use induction on $r$. For $r=1$ the statement is obvious (with $I_0={\bf 0}$).
Suppose the statement holds for $r-1$, so that there is an exact sequence:
\be
\label{foo_seq1}
0 \longrightarrow I_{r-2} \stackrel{j}{\longrightarrow} F'_{r-1} \stackrel{p}
{\longrightarrow} O((r-1)p) \longrightarrow 0
\ee
Let $s_1...s_{r-1}$ be a basis of $H^0(I_{r-2})$ and $s_{r-1}$ a section of 
$ F'_{r-1}$ such that $s_1...s_{r-1}$ is a basis of $H^0(F'_{r-1})$ and 
(using \ref{short}) such that $p_*(s_{r-1}) \in H^0(O(p))-\{0\}$.
Then $s_1(e)...s_{r-2}(e)$ are linearly independent for all $e \in E$
By Proposition \ref{ker_prop}, we have that $s_{r-1}(e) \in F'_{r-1,e} -
<s_1(e)...s_{r-2}(e)>=F'_{r-1,e}-I_{r-1,e}$ for all $e \neq p$. 
Now use the recursive definition (\ref{twisted_1}) of $F'_r$. 
This shows that we can choose $s_r \in H^0(F'_r)$ such that 
$s_1...s_r$ is a basis of $H^0(F'_r)$ and such that the induced section 
${\overline s}_r \in H^0(O(p))$ has zero divisor $({\overline s}_r)=(p)$.
Since $s_{r-1}(p)=0$, Proposition \ref{ker_prop} applied to $F'_r$ shows that 
$s_1(e)...s_{r-2}(e),s_r(e)$ are linearly independent for {\em all} $e \in E$,
while $s_1(e)...s_{r-2}(e),s_{r-1}(e)$,\\ $s_r(e)$ are linearly independent 
for 
$e \neq p$. Thus $s_1(e)...s_{r-2}(e),s_r(e)$ determine a trivial subbundle 
$I_{r-1}$ of $F'_r$ and $s_{r-1}(e)$ belongs to this subbundle iff $e=p$  
(where $s_{r-1}(p)=0$). 
It follows that the induced section ${\overline s}_{r-1}$ of the line bundle 
$L:=F'_r/I_{r-1}$ vanishes only at $p$. Since $\deg F'_r=r$, we have 
$\deg L=r$ 
so that $\deg ({\overline s}_{r-1})=r$. Therefore 
$({\overline s}_{r-1})=rp$ and $L\approx O(rp)$. This gives an exact sequence :
\be
\label{foo_seq2}
0 \longrightarrow I_{r-1} \stackrel{j}{\longrightarrow} F'_r \stackrel{p}
{\longrightarrow} O(rp) \longrightarrow 0
\ee

{\em Show that $O(q)\otimes F_r$ for $q \in T^{(p)}_r(E)$ are also 
extensions of $O(rp)$ by $I_{r-1}$}

Since $q \in T^{(p)}_r(E)$, we have $O(rq)\approx O(rp)$. Combined with 
(\ref{foo_seq2})(applied for $p$ substituted with $q$), 
this gives the desired statement.

{\em Show that any indecomposable $A'$ which can be presented as such an 
extension is of this form}

If $A'$ is an extension of $O(rp)$ by $I_{r-1}$, then 
$\det A'\approx O(rp)$. If $A'$ is indecomposable then $A:=A'\otimes O(-p)$
belongs to ${\cal E}(r,0)$, so that 
$A\approx O(q-p)\otimes F_r$ for some $q \in E$.
($q$ is uniquely determined by $A$). Then $A'\approx O(q)\otimes F_r$, 
so that $\det A'\approx O(rq)$. Thus we must have $O(rq)\approx O(rp)$
i.e. $q \in T_r^{(p)}(E)$. This finishes the proof. $\Box$

\begin{Theorem}
\label{F_criterion}
	Let $V$ be a degree zero holomorphic vector bundle of rank $r$ over $E$
and let $V':=V\otimes O(p)$. The following statements are equivalent :

(a) $V$ is holomorphically equivalent to $O(q)\otimes F_r$, where $q$ is a 
point of $E$

(b) There exists a $\C$-basis $(s_1 ... s_r)$ of $H^0(V')$ with the following 
properties :

	(b1) $s_1(e) ... s_r(e)$ is a basis of $V'_e$ for all 
	$e \in E - \{ p \}$

	(b2) $s_1(p)=0$ and 
	$s_2(p),...,s_r(p)$ are linearly independent in $V'_p$ 

	(b3) $\deg s_1\wedge s_2\wedge ... \wedge s_j (p)= j$ for all $j=1..r$.

(c) The following conditions are satisfied :

(c1) $h^0(V')=r$

(c2) ${\cal Z}(V')=\{p\}$

(c3) There exists a {\em nondegenerate} filtration 

\be
{\cal K}: 0=K_0\subset K_1 \subset ... \subset K_r:=H^0(V')
\ee
of $H^0(V')$, with associated filtration 
\be
0 =W_0\subset W_1 \subset ... \subset  W_r:=V' 
\ee
of $V'$, having the properties :

(c31) $K_j=\{s \in H^0(V') | s(p) \in (W_{j-1})_p\}$ 
(i.e. $K_j=\phi_p^{-1}(W_{j-1})$), $\forall j=1..r$

(c32) $\delta^{\cal K}_j(p)=j, \forall j=1..r$

Moreover, in this case we have $W_j\approx F'_j$ and 
$K_j=H^0(W_j)\approx H^0(F'_j)$ for all $j=1..r$.

\end{Theorem}

Note that $s_2...s_r$ generate a trivial subbundle $I_{r-1}$ of $F'_r$ (since 
they are everywhere linearly independent), while the section $s_1$ is incident
on $I_{r-1}$ at $p$ in order $r$. This is in agreement with the previous 
proposition. The precise manner of incidence of $s_1$ on $I_{r-1}$ is 
controlled by condition $(b3)$.

Note that $(c31)$ acts as an inductive definition of the filtration ${\cal K}$.
For $j=1$, $(c31)$ gives $K_1=\ker \phi_p=K_p(V')$. The map 
$\psi_1:\Gr^1(H^0(V'))\rightarrow \Gr^1(V')$ gives the subbundle 
$W_1=\psi_1(K_1)$. Then $(c32)$ for $j=2$ defines $K_2$, the map $\psi_2$ 
gives $W_2=\psi_2(K_2)$ and so on. 
In particular, ${\cal K}$ is naturally associated to $F'_r$
\footnote{Of course, $F'_r$ are only determined up to isomorphism. Naturality 
heer means that such an isomorphism is compatible with the filtrations 
${\cal K}$}. 
It is easy to see from the proof of the theorem below that ${\cal K}$ is 
nothing 
other then the cohomology filtration induced by the standard Jordan-Holder  
filtration of $F'_r$ :
\be
\label{bundle_filtration}
0\longrightarrow F'_1\longrightarrow F'_2 \longrightarrow ...\longrightarrow
F'_{r-1}\longrightarrow F'_r 
\ee
Indeed, (\ref{bundle_filtration}) has the  partial sequences:
\be
0\longrightarrow F'_{j-1}\longrightarrow F'_j \longrightarrow O(p)
\longrightarrow 0
\ee
(for $j=2...r$). Since $H^1(F'_{j-1})=0, \forall j=2..r$, these give the 
exact sequences :
\be
0\longrightarrow H^0(F'_{j-1})\longrightarrow H^0(F'_j) \longrightarrow 
H^0(O(p))\longrightarrow 0
\ee
which combine to give the filtration :
\be
0\longrightarrow H^0(F'_1)\longrightarrow H^0(F'_2) \longrightarrow ... 
\longrightarrow H^0(F'_{r-1})\longrightarrow H^0(F'_r) 
\ee
of $H^0(F'_r)$. This can be identified with the filtration ${\cal K}$ in the 
theorem.

{\em Proof:}

{\em Show that (a) implies  (b) }

We proceed by induction on $r$. For $r=1$, the statement is 
trivial.
Let $r\geq 2$ and suppose the statement holds for $r-1$. 
By the above discussion, we can choose bases 
$\sigma_1...\sigma_{r-1}$ of $H^0(F'_{r-1})$, 
$\sigma_r$ of $H^0(O(p))$ and $s_1...s_r$ of $H^0(F'_r)$ such that 
$j_*(\sigma_1)=s_1...j_*(\sigma_{r-1})=s_{r-1}$ and $p_*(s_r)=\sigma_r$.

Since the result holds for $r-1$, we can further assume that 
$\sigma_1...\sigma_{r-1}$ satisfy the properties $(b)$ for $r$ replaced with 
$r-1$. Since $p_*(s_r)=\sigma_r$ and $(\sigma_r)=p$, it folows that 
$s_r(e) \in (F'_r)_e-j_e((F'_{r-1})_e), \forall e \in E -\{p\}$, while 
$s_r(p) \in j_p((F'_{r-1})_p)$. Since $j_e$ is injective for all $e \in E$, 
and since $\sigma_1 ... \sigma_{r-1}$ satisfy $(b1)$, we see that 
$s_1(e)... s_r(e)$ are linearly independent for all $ e \in E-\{p\}$, so 
that $s_1 ... s_r$ satisfy $(b1)$.
By $(b2)$ for $\sigma_1...\sigma_{r-1}$ we obtain that $s_1(p)=0$ 
and $s_2(p)...s_{r-1}(p)$ are linearly independent.

	Now suppose that $s_r(p) \in <s_2(p) ... s_{r-1}(p)>$. Then 
$s_r(p)=\alpha_2 s_2(p) +.. + \alpha_{r-1}s_{r-1}(p)$. Then 
$s:=s_r -\alpha_2 s_2 -.. - \alpha_{r-1}s_{r-1}$ is a regular section 
of $F'_r$ which vanishes at $p$. Since $s_r$ is linearly independent of 
$s_1...s_{r-1}$, it is clear that $s$ is linearly independent of 
$s_1...s_{r-1}$. In particular, $s$ is linearly independent of $s_1$.
This implies that we have two linearly independent sections $s_1$, $s$ of 
$F'_r$, both vanishing at $p$. Since this is impossible by virtue of 
Proposition \ref{ker_prop}, it follows that $s_2(p) ... s_r(p)$ are 
linearly independent and $(b2)$ holds.

	Since $p_*(s_r)=\sigma_r$ has a simple zero at $p$, it follows 
that $s_r$ vanishes in order $1$ along the subbundle 
$j_*(F'_{r-1})$ of $F'_r$. Since $(b3)$ holds for $F'_{r-1}$ by the induction 
hypothesis, we also know that  $s_j$ vanishes in order $1$ along the 
subbundle $W_j$ of $F'_{r-1}$, where $W_j=W_{s_1...s_j}$, for all $j=1..r-1$.
In particular, we have $s_j(e)=z{\tilde s}_j(p) + s_{0j}(e)$, with 
$s_{0j} \in H^0(W_{j-1})$ 
for all $j=1..r-1$, and all $e$ sufficiently close to $p$. This implies that 
$s_1(e) \wedge ... \wedge s_{r-1}(e) = z^{r-1}{\tilde s}_1(e) \wedge ... 
\wedge {\tilde s}_{r-1}(e)$ so that  
${\tilde s}_1(e) \wedge ...\wedge {\tilde s}_{r-1}(e) \neq 
0$ for $e$ near $p$. This shows that ${\tilde s}_1, ..., {\tilde s}_{r-1}$ 
give a local holomorphic frame of $F'_{r-1}$ in a vicinity of $p$. 
Then by Proposition \ref{frame_criterion}, we must have 
$\deg {\tilde s}_1\wedge ...\wedge {\tilde s}_{r-1}\wedge s_r (p)=1$,
so that $\deg s_1\wedge ... \wedge s_r (p)=r$. Thus $(b3)$ 
holds for $F'_r$. 
Thus $(a)$ implies $(b)$.

{\em Show that (b) implies (c)}

Assume $(b)$ holds. Then $(c1)$ and $(c2)$ are obvious.
We can construct a filtration:
\be
{\cal K}: 0:=K_0\subset K_1:=<s_1> \subset K_2:=<s_1,s_2> \subset ...\subset
K_r:=H^0(V')
\ee
of $H^0(V')$, and an associated filtration :
\be
{\cal W}: 0:=W_0\subset W_1 \subset W_2 \subset ...\subset W_r:=V'
\ee
of $V'$, as explained in the previous subsection. 
Let us analyze the situation at the point $p$.

{\em Claim:} For each $j=1..r$, we have $\deg _{W_{j-1}}s_j(p)=1$ 
and $s_2(p)...s_j(p)$ is a $\C$-basis of $(W_{j-1})_p$.

We prove the claim by induction on $j$. For $j=1$ we have 
$W_{j-1}=W_0={\bf 0}$ and, by $(b3)$, we have $\deg _{W_0}s_1(p)=\deg s_1 (p)=1$.
The second part of the claim is trivial in this case.

Now let $j \in \{2...r\}$ and assume that the claim is true for all $j'<j$. 
Fix a local coordinate $z$ on $E$, centered at $p$. 
By Proposition \ref{osc_vector}, we can write :
\be
s_k(e)=z{\tilde s}_k(e) + s_{0k}(e), ~\forall k=1..j-1
\ee
for all $e$ sufficiently close to $p$, where 
${\tilde s}_k(p) \in V'_p-(W_{k-1})_p$ and $s_{0k}$ is a local section of 
$W_{k-1}$. Then $s_1(e)\wedge ...\wedge s_{j-1}(e)=z^{j-1}{\tilde s}_1(e)
\wedge ... \wedge {\tilde s}_{j-1}(e)$ for $e$ close to $p$. By $(b3)$, we 
have ${\tilde s}_1(p)\wedge ... \wedge {\tilde s}_{j-1}(p) \neq 0$ and by 
continuity ${\tilde s}_1(e)\wedge ... \wedge {\tilde s}_{j-1}(e) \neq 0$ 
for $e$ close to $p$. Thus ${\tilde s}_1 ...  {\tilde s}_{j-1}$ is a 
local holomorphic frame of $W_{j-1}$ around $p$. 
We obtain :
\be
\nonumber
s_1(e)\wedge ... \wedge s_j(e)=z^{j-1}{\tilde s}_1(e)\wedge ... \wedge 
{\tilde s}_{j-1}(e)\wedge s_j(e)
\ee
(for $e$ close to $p$), which together with $(b3)$ gives :
\be
\nonumber
\label{foo}
\deg {\tilde s}_1\wedge ... \wedge {\tilde s}_{j-1}\wedge s_j (p) = 1
\ee
Since ${\tilde s}_1 ...  {\tilde s}_{j-1}$ is a 
local holomorphic frame of $W_{j-1}$ around $p$, this shows, by Proposition
 \ref{frame_criterion}, that $\deg _{W_{j-1}}s_j(p)=1$.

Since 
${\tilde s}_1(p)\wedge ... \wedge {\tilde s}_{j-1}(p)\wedge s_j(p)=0$ 
by (\ref{foo}), it follows that 
$s_j(p)\in <{\tilde s}_1(p) ... {\tilde s}_{j-1}(p)>$ \\ $=(W_{j-1})_p$. 
By the induction hypothesis, $s_2(p)...s_{j-1}(p)$ is a basis of 
$(W_{j-2})_p \subset (W_{j-1})_p$,
so that $s_2(p) .... s_{j-1}(p) \in (W_{j-1})_p$. 
Thus, the vectors $s_2(p)....s_j(p)$ all belong to the $j$-dimensional 
vector space $(W_{j-1})_p$. Since they are linearly independent by $(b2)$, 
they must form a basis of this subspace. This finishes the proof of the claim.

Since 
$\delta^{\cal K}_j(p)-\delta^{\cal K}_{j-1}(p)=\deg _{W_{j-1}}s_j(p)$, 
the first part of the claim implies $(c32)$.
The second part of the claim is easily seen to imply $(c31)$.
Thus $(b)$ implies $(c)$.

{\em Show that (c) implies (a)}

Again proceed by induction on $r$. For $r=1$ the statement is immediate.

Now let $r >1 $ and suppose that $(c)\Rightarrow (a)$ holds for $r-1$. 
Also assume that $V'$ satisfies $(c)$. 
Since ${\cal K}$ is nondegenerate, we have $\dim_{\C}K_j=j$ for all $j=1..r$. 
In particular, $K_1$ is a line bundle. By $(c31)$ and $(c31)$ we have 
$K_1\approx O(p)$. Define $W':=V'/K_1$. We have an exact sequence:
\be
\label{s_1}
0 \longrightarrow W_1\stackrel{j}{\longrightarrow}V'\stackrel{p}{\longrightarrow} W'
\longrightarrow 0
\ee
To show $(a)$ it suffices to show that $W'\approx F'_{r-1}$ and that 
(\ref{s_1}) is nonsplit. By the induction hypothesis, 
to show $W'\approx F'_{r-1}$ it suffices to show that $W'$ satisfies $(c)$ 
for $r-1$. We proceed to do this. 

{\em Show that $W'$ satisfies $(c1)$.} 
Since $H^1(O(p))=0$, (\ref{s_1}) gives :
\be
\label{s_2}
0 \longrightarrow H^0(O(p))\stackrel{j_*}{\longrightarrow}H^0(V')
\stackrel{p_*}{\longrightarrow} H^0(W') \longrightarrow 0
\ee
Thus $h^0(W')=r-1$. 

{\em Show that $W'$ satisfies $(c2)$.} 
For each $e \in E -\{p\}$ we have a commutative diagram with exact rows:
\be
\begin{array}{ccccccccc}
0 & \longrightarrow & H^0(K_1) & \stackrel{j_*}\longrightarrow &H^0(V') 
&\stackrel{p_*}{\longrightarrow} & H^0(W') & \ \longrightarrow 0 \\
\ &               \ & \phi^{O(p)}_e \downarrow &       \ & \phi_e \downarrow 
&              \ & \phi'_e \downarrow & \  \\
0 & \longrightarrow & K_{1,e} & \stackrel{j_e}{\longrightarrow} & V'_e & 
\stackrel{p_e}{\longrightarrow} & 
W'_e & \longrightarrow & 0 \\
\end{array}
\ee
where the vertical arrows represent the evaluation maps. 
$\phi^{O(p)}_e$ is trivially an isomorphism, while $\phi_e$ is 
an isomorphism since $V'$ satisfies $(c1)$ and 
$(c2)$. Thus $\phi'_e$ is an isomorphism. We will see below that $\phi_p$ 
is not injective. Thus $W'$ satisfies $(c2)$.

{\em Show that $W'$ satisfies $(c31)$ and $(c32)$.}
First we show that $K_j=H^0(W_j)$ for all $j=1..r$. To see this, note that 
$(c31)$ implies $H^0(W_{j-1}) \subset K_j$ for all $j$. This inclusion is 
{\em strict} (otherwise $\phi_e|_{K_j}$ for $e \neq p$ would coincide with the 
evaluation map of $W_{j-1}$; since $\phi_e$ is injective and $\dim_{\C}K_j=j$, 
this would contradict the rank theorem). We also trivially have 
$K_j \subset H^0(W_j)$ for all $j$. This gives 
$H^0(W_{j-1}) \subset K_j\stackrel{\neq}{\subset}H^0(W_j)$ for all $j$ 
and since $\dim_{\C}K_j=j$ we obtain $K_j=H^0(W_j)$.

${\cal K}$ induces a filtration ${\cal K}'$: 
\be
0=K'_0\subset K'_1\subset ...\subset K'_{r-1}
\ee
by $K'_j:=p_*(K_{j+1})$ for all $j=1..r-1$. By (\ref{s_2}) we have 
$K'_{r-1}=H^0(W')$ and $\dim_{C}K'_j=j$ for all $j=1..r-1$. On the other hand, 
the filtration ${\cal W}$ of $V'$ induces a nondegenerate 
filtration ${\cal W}'$ of $W'$:
\be
0=W'_0\subset W'_1\subset ...\subset W'_{r-1}=W'
\ee
by $W'_j:=p(W'_{j+1})$. 

For each $j=1..r-1$ we have a commutative diagram:
\be
\begin{array}{ccccccccc}
0 & \longrightarrow & K_1 & \stackrel{j_*}{\longrightarrow} & K_{j+1} 
&\stackrel{p_*}{\longrightarrow}& K'_j & \longrightarrow 0 \\
\ &    \ & \phi_p \downarrow &       \ & \phi_p \downarrow 
&     \ & \phi'_p \downarrow & \  \\
0 & \longrightarrow & K_{1,p} & 
\stackrel{j_p}{\longrightarrow} & W_{j,p} & 
\stackrel{p_p}{\longrightarrow} & 
W'_{j-1,p} & \longrightarrow & 0 \\
\end{array}
\ee
(we have $\phi'_p(K'_j)=\phi'_p(p_*(K_{j+1}))=p_p(\phi_p(K_{j+1}))\subset 
p_p(W_{j,p})=W'_{j-1,p}$ where we used $(c31)$ for $V'$ ).
Commutativity of the second square gives $p_*^{-1}(\phi_p^{' -1}(W'_{j-1,p}))=
\phi_p^{-1}(p_p^{-1}(W'_{j-1,p}))=\phi_p^{-1}(W_{j,p})=K_{j+1}$ where we used 
$(c31)$ for $V'$. Thus $\phi_p^{-1}(W'_{j-1,p})$\\$=
p_*(K_{j+1})=K'_j$ and ${\cal K}',{\cal W}'$ satisfy $(c31)$. 

Now pick $s_j \in K_j-K_{j-1}$ for all $j=1..r$ and let 
$\sigma_j:=p_*(s_{j+1})$ for all $j=1..r-1$. Then 
$\sigma_j \in K'_j -K'_{j-1}$ and $\deg _{W'_{j-1}}\sigma_j (p)=\deg _{W_j} 
s_{j+1}(p)$ by Proposition \ref{induced_section}. Using $(c32)$ for $V'$ 
and $\delta^{\cal K}_j(p)-\delta^{\cal K}_{j-1}(p)=\deg_{W_{j-1}}\sigma_j(p)$,
this immediately implies $(c32)$ for $W'$. (In particular, we have 
$\ker \phi'_p=K'_1 \neq 0$, as announced above). 

Now suppose that (\ref{s_1}) is split. Then 
$V' \approx O(p) \oplus F'_{r-1}$. Since $O(p)$ and 
$F'_{r-1}$ both posess nonzero 
sections which vanish at $p$, 
this immediately gives two linearly independent sections of $V'$ which vanish
at $p$. But $(c3)$ implies $d_p(V')=1$, which gives a 
contradiction. Thus, (\ref{s_1}) cannot split and we must have 
$V'\approx F'_r$ 
and $V\approx F_r$. Thus $(c)$ implies $(a)$. 

To prove the last statement of the theorem it suffices to note that 
each of the bundles $W_j$ in $(c)$ also satisfies $(c)$ for the 
appropriate rank. $\Box$

It is now possible to analyze the freedom in the choice of ${\tilde s}_j$ 
and define a notion of canonical bases of $H^0(F'_r)$ by imposing further 
conditions on $s_1..s_r$. This leads to a concrete description of the 
endomorphisms of $F'_r$ via their induced action on $H^0(F'_r)$, which  
can then be used to  analyze the endomorphisms of a general 
degree zero semistable  
bundle by using the results of the next subsection. 
Since this is not directly related to the main focus of the present paper, we 
will not proceed down that path.

\subsection{The main theorem}

The results of the previous subsection immediately lead to:

\begin{Theorem}
\label{final_theorem}

Let $V$ be a degree zero holomorphic vector bundle of rank $r$ over $E$ and 
$V'=V \otimes O(p)$. Let $\phi_e$ be the evaluation map of $V'$ and 
$S:={\cal Z}(V'):=\{t \in E | K_t(V') \neq 0\}$. 
The following are equivalent :

(a) $V$ is semistable 

(b0) $h^0(V')=r$ 

(b1) The set $S$ is finite. Let 
$d_t:=\dim_{\C}K_t(V')$ for all $t \in S$.

(b3) There exists a direct sum decomposition :
\be
H^0(V')=\oplus_{t \in S}{\oplus_{i=1..d_t}{K^{(i)}_{r_{t,i}}(t)}}
\ee
with $\dim_{\C}K^{(i)}_{r_{t,i}}(t)=r_{t,i}$ and {\em nondegenerate} 
filtrations :
\be
{\cal K}^{(i)}(t) \ : \ 
0=K^{(i)}_0(t)\subset K^{(i)}_1(t)\subset ...\subset K^{(i)}_{r_{t,i}}(t)
\ee
with $\psi$-associated bundle filtrations: 
\be
{\cal W}^{(i)}(t) \ : \ 
0=W^{(i)}_0(t)\subset W^{(i)}_1(t)\subset ...\subset W^{(i)}_{r_{t,i}}(t)
\ee

with the properties :

(b31) We have 
$V'_t=R_t(V')\oplus \oplus_{i=1..d_t}{(W^{(i)}_{r_{t,i}}(t))_t}$, for all  
$t \in S$.

(b32) $\delta^{{\cal K}^{(i)}(t)}_s(t)=s$ for all $t \in S$, all $i=1..d_t$
and all $s=1..r_{t,i}$

(b33) The induced filtrations 
$0=\phi_t(K^{(i)}_1(t))\subset ...\subset \phi_t(K^{(i)}_{r_{t,i}}(t))$
in $V'_t$ are nondegenerate for all $t\in S$ and $i=1..d_t$

(c) The following conditions are satisfied:

(c1) $h^0(V')=r$

(c2) The set $S$ is finite. Let $d_t=\dim_{\C} 
K_t(V')$, $\forall t \in S$

(c3) There exists a basis $(s^{(i)}_{t,j})_{t \in S, i=1..d_t, j=1..r_{t,i}}$ 
of $H^0(V')$ ($\sum_{t \in S, i=1..d_t}{r_{t,i}}=r$) with the properties :

(c31) $\deg (\Lambda_{i=1..d_t, t' \in S}
{s^{(i)}_{t',1}\wedge...\wedge s^{(i)}_{t',r_{t',i}}})(t)=
\sum_{i=1..d_t}{r_{t,i}}$,  
$\forall t \in S$

(c32) $(s^{(i)}_{t,j})_{j=2...r_{t,i}}$ are 
linearly independent for all $t\in S$ and all $i=1..d_t$.

(c33) $\deg (s^{(i)}_{t,1}\wedge...\wedge s^{(i)}_{t,j})(t)=j$, 
$\forall t \in S, \ \forall i=1..d_t, \ \forall j=1..r_{t,i}$

In this case, we have:

$V' \approx \oplus_{t \in S}{\oplus_{i=1 ... d_t}{O(t)\otimes F_{r_{t,i}}}}$

\end{Theorem}

The proof should be rather obvious by now. Instead of writing down all of its 
details, let us try to make the statement of the theorem 
look less formidable. Clearly the 
bundles $W^{(i)}_{r_{t,i}}(t)$ are isomorphic to $O(t)\otimes F_{r_{t,i}}$,
while $W^{(i)}_j(t)\approx O(t)\otimes F_j$ give their canonical filtrations. 
$d_t$ is the number of different indecomposable bundles which  
multiply $O(t)$ in the decomposition of $V'$. These bundles are just 
$W^{(i)}_{r_{t,i}}(t)$, and have ranks $r_{t,i}$ (of which some may coincide). 
Conditions $(b32)$ and $(b33)$ or, equivalently, conditions $(c32)$   
and $(c33)$ are needed to assure that 
$W^{(i)}_{r_t,i}(t)\approx F_{r_{t,i}}$. Conditions $(b31)$, 
respectively $(c31)$ are needed in order to have a {\em direct} factor of 
the form  
$O(t)\otimes \oplus_{i=1..d_t}{F_{r_{t,i}}}$ in the decomposition of $V'$. 

Note that the spectral divisor is: 
\be
\Sigma_V=\sum_{t \in S}{\sum_{i=1..d_t}{r_{t,i}~t}}
\ee
We immediately obtain
\footnote{This result can also be obtained without making use of Theorem  
\ref{final_theorem}}:

\begin{Corrolary}
\label{spectral_div}
Let $V$ be a degree zero semistable holomorphic vector bundle over $E$ and 
$V'=V \otimes O(p)$. Let $s_1...s_r$ be a $\C$-basis of $H^0(V')$.
Then the spectral divisor of $V$ is given by :
\be
\label{sp_div}
\Sigma_V=(s_1\wedge ... \wedge s_r)
\ee
\end{Corrolary}
{\em Proof:} Since $(s_1\wedge ... \wedge s_r)$ is independent of the choice 
of the basis of sections $s_1...s_r$, we can choose $s_1..s_r$ to have 
the properties listed in $(c)$ of Theorem \ref{final_theorem}. Then the 
conclusion is obvious.
$\Box$

This shows that the spectral divisor can be computed by an obvious 
adaptation of the methods of \cite{CGL} even in the general case. 
However, the divisor $(s_1\wedge...\wedge s_r)$ alone cannot give us enough 
information to test 
semistability and/or determine the splitting type.

Starting from the above theorem, it is relatively straightforward to develop 
an algorithm for testing semistability of $V$ and determining its splitting 
type by doing a series of simple manipulations on an arbitrary basis of 
$H^0(V')$. Instead of presenting the algorithm in its full generality (which 
requires introducing a slightly tedious amount of notation), we will 
show explicitly how this can be implemented in the simpler case when one is 
interested in identifying degree zero {\em fully decomposable} semistable 
bundles. This is explained in section 3 below. 

\subsection{The spectral divisor in the monad case and a `moduli problem'}

In this subsection we consider the case when $V$ is given by the cohomology
of a monad:

\be
\label{monad}
0 \longrightarrow \oplus_{j=1..s}{O_E}\stackrel{f}{\longrightarrow}
\oplus_{a=1..m}{O(D_a)}\stackrel{g}{\longrightarrow}O(D_0)\longrightarrow 0
\ee	
Here $D_a,D_0$ are some divisors on $E$.
We define the twisted bundles and exact  sequences as before.
We denote all twisted objects by a prime. As usual, we twist by $O(p)$ with $p$ 
an arbitrary point on $E$.$p$ is fixed throughout the following discussion.
We have $m=r+s+1$ where $r:=\rank V$.

Write (\ref{monad}) as the pair of exact sequences :

\be
\label{sequence1}
0 \longrightarrow \ker g \hookrightarrow \oplus_{a=1..m}{O(D_a)}\stackrel{g}
{\longrightarrow}O(D_0)\longrightarrow 0
\ee

\be
\label{sequence2}
0 \longrightarrow\oplus_{j=1..s}{O_E}\stackrel{f}{\longrightarrow}\ker g
\stackrel{p}{\longrightarrow}V{\longrightarrow}0
\ee

By taking degrees we obtain :
\be
\label{degrees}
\deg V=\sum_{a=1..m}{\deg D_a}-\deg D_0=\deg (\ker g)
\ee

We have:
\begin{Proposition}
\label{ker_ss}
The following are equivalent :

(a) $V$ is semistable and of degree zero

(b) $\ker g$ is semistable and of degree zero

\end{Proposition}

Proof:

Assume that $(a)$ holds. Then the sequence (\ref{sequence2}) shows that 
$\ker g$ is an extension of $\oplus_{j=1..s}{O_E}$ by $V$.
As both these bundles are semistable and of slope zero, a standard result of 
Seshadri (see. for example, \cite{Seshadri}) immediately entails $(b)$.

Assume $(b)$ holds. Then (\ref{sequence2}) shows that $V=\coker f$
and since $\oplus_{j=1..s}{O_E}$ and $\ker g$ are both semistable and of slope 
zero we can use another result of Seshadri to obtain $(a)$. 
$\Box$

This proposition reduces the study of semistabilty of $V$ to that 
of $\ker g$. In particular, we see that 
semistability of $V$ depends only on the properties of the map $g$ and on
the bundles $\oplus_{a=1..m}{O(D_a)}$ and $O(D_0)$.

For the following we assume that $\oplus_{a=1..m}{\deg D_a}=\deg D_0:=d$. 
with $d \ge 0$. We let $d_a:=\deg D_a$. 
Then (\ref{degrees}) assures us that $\deg V=\deg \ker g=0$.

Now {\em suppose that $V$ is semistable} .

Then by Proposition \ref{ker_ss} $\ker g$ is also semistable .
Then Lemma \ref{cohom_dim} assures us that $H^1(\ker g')=0$.
Noting that $H^1(O(p))$ also vanishes by the Riemann-Roch theorem, it follows
that by twisting the two exact sequences above and taking cohomology we obtain
two {\em short} exact sequences :

\be
\label{cohomology1}
0 \longrightarrow H^0(\ker g') \hookrightarrow \oplus_{a=1..m}{H^0(O(D'_a))}
\stackrel{g_*}{\longrightarrow}H^0(O(D'_0))\longrightarrow 0
\ee

\be
\label{cohomology2}
0\longrightarrow\oplus_{j=1..s}{H^0(O(p))}\stackrel{f_*}{\longrightarrow}
H^0(\ker g')\stackrel{p_*}{\longrightarrow}H^0(V'){\longrightarrow}0
\ee
where $D'_a:=D_a+p,D'_0:=D_0+p$ and we denoted $f\otimes id, g\otimes id$ by 
the same letters for simplicity.
The collapse of the cohomology sequence associated to 
(\ref{sequence1}) is a direct consequence of the semistability of $\ker g$.

Since $d+1$ is positive, the Riemann-Roch theorem tells us that 
$h^0(O(D'_0))=\deg (D'_0)=\deg D_0+1=d+1$. Since $\ker g$ is semistable 
and of degree zero, Lemma \ref{cohom_dim} gives $h^0(\ker g')=
\rank (\ker g)=r+s=m-1$; then (\ref{cohomology1}) gives 
$h^0(\oplus_{a=1..m}O(D'_a))=m+d$. This last fact is not a consequence of 
Riemann-Roch unless $d_a$ are all nonnegative.

\begin{Proposition}
\label{cover_rel}
Let $\Sigma_{\ker g}$ and $\Sigma_V$ be the spectral divisors of $\ker g$, 
respectively $V$.
Then $\Sigma_{\ker g}=\Sigma_V+sp$.
\end{Proposition}

{\em Proof:}
Since (\ref{cohomology2}) is an exact sequence of 
vector spaces, it must split. We can thus choose a 
basis $v_1 ... v_{r+s}$ of 
$\ker g'$ with the properties :

(1)$v_1=f_*(w_1)...v_s=f_*(w_s)$, where $w_1...w_s$ is a 
basis of $A:=\oplus_{j=1..s}{H^0(O(p))}$

(2)$p_*(v_{s+1}):=u_1...p_*(v_{s+r}):=u_r$ is a basis of $H^0(V')$

The canonical isomorphism $\det (\ker g') \approx \det (A)\otimes \det (V')$
maps the section $v_1\wedge ..\wedge v_{r+s}\in H^0(\det (\ker g'))$ into 
the the section $(w_1\wedge...\wedge w_s)\otimes (u_1\wedge ...\wedge u_r)
\in H^0(\det A)\otimes H^0(\det V')\subset H^0(\det A \otimes \det V')$. 
Thus:

\be
\nonumber
\Sigma_{\ker g}=(v_1\wedge...\wedge v_{r+s})=
(w_1\wedge...\wedge w_s\otimes u_1\wedge ...\wedge u_r)
=(w_1\wedge...\wedge w_s) + (u_1\wedge ...\wedge u_r)=
sp+\Sigma_V
\ee  
where in the first and last line we used the corrolary to 
Theorem \ref{final_theorem}. 
$\Box$

The relation between $\Sigma_{\ker g}$ and the bundle 
$B:=\oplus_{a=1..m}{O(D'_a)}$ is more complicated. The reason is that 
there is no simple connection between the local behaviour of the sections 
of $\ker g$ and the sections of $B$\footnote{This happens because typically 
we have $d_a > 0$ for some $a$. Then $O(D'_a)$ is quasi-ample (the evaluation 
map is surjective everywhere), which  to complications.}.
To extract more information about $\ker g$, one has to undertake 
a more 
detailed study based on the properties of the map $g$. In particular, 
one would like to find necessary and sufficient conditions on $g$ such that 
$\ker g$ is semistable and describe the associated moduli space of $g$. 
Although we will not 
attempt this here, let us formulate the geometric set-up of the problem.

Theorem \ref{final_theorem} reduces the semistability condition  for 
$\ker g$ to conditions on the subspace $W:=H^0(\ker g')$ of $U:=H^0(B)$. 
We have $\dim_{\C}U=m+d$ and semistability requires that $\dim_{\C}W=m-1$. 
Let $H_e:=\ker \phi^B_e$, for all $ e\in E$. Note that 
$\phi_e^{\ker g'}=\phi_e^{B}|_W$, so that 
$\ker \phi_e^{\ker g'}=H_e \cap W$. 

Suppose  for simplicity that all $D_a$ are effective and that 
$\Card\{a \in \{1..m\} | d_a=0\}=\nu$. Since $\phi_e^B=\oplus_{a=1..m}
{\phi_e^{O(D'_a)}}$ for all $e \in E$, we have $\ker (\phi_e^B)=
\oplus_{a=1..m}{\ker \phi_e^{O(D'_a)}}$. 
With our assumptions, we have $\codim_{\C}(\ker \phi^{O(D'_a)}_e)=1$
for all $a$ with $d_a>0$ and all $e \in E$, while for all $a$ with $d_a=0$ 
(i.e. $D_a=0,~ O(D'_a)=O(p)$) we have  
$\codim_{\C}(\ker \phi^{O(D'_a)}_e)=1$ for $e \neq p$ and  
$\codim_{\C}(\ker \phi^{O(D'_a)}_p)=0$. 
Thus $\codim_{\C}H_e=m$ for all $e \neq p$ while $\codim_{\C}H_p=m-\nu$. 

Note that $\dim_{C}W+\dim_{\C}H_e = \dim_{\C}U-1$ for $e \neq p$ while 
$\dim_{\C}W+\dim{\C}H_p =\dim_{\C}U +\nu-1$. For given divisors $D_a$ and a given 
map $g$, $W \cap H_e$ will have fixed dimension $D$ for almost all points 
$e \in E$. The points where the dimension of this intersection increases 
correspond to the points of the set ${\cal Z}(\ker g')$.

If $\nu > 1$, it follows that $\dim_{\C} W \cap H_p \ge \nu-1$. 
On the other hand, we cannot deduce any simple 
lower bound on 
$\dim_{\C}W \cap H_e$ for $e \neq p$. 

Geometrically, we are given a map $H:E \rightarrow Sbsp(U)$, 
$H(e):=H_e, \forall e \in E$
from $E$ to the set of subspaces of the $m+d$-dimensional $\C$-vector space 
$U$. The precise form of this map is completely fixed by the bundle $B$.
As $e$ varies in $E$, $H_e$ describes a complicated trajectory in $Sbsp(U)$. 
Generically on $E$, $H_e$ has codimension $m$, except at the point $e=p$ where
it has codimension $m-\nu$. 

Giving a semistable subbundle of $B$ of the form $\ker g$ requires 
giving the $m-1$ dimensional subspace $W$ of $U$, with the property that 
it is complementary to $H_e$ for a generic $e\in E$ and satisfying the 
other conditions in Theorem \ref{final_theorem}. The precise position of 
$W$ inside $U$ is controlled by the map $g$. 
It is not hard to see that the remaining 
conditions in the theorem can be expressed in terms of 
`incidence relations' constraining the `speeed of incidence' of $W$ on $H_e$
as $e \rightarrow t_i$; this is similar to the discussion of Section 2.

The set-up above allows us to reduce the problem of determining the maps 
$g$ giving 
a semistable $\ker g$ to a problem 
in linear algebra and analysis. In particular, it is ideal for 
extracting information about `moduli'.
The `trajectory' of $H_e$ is, however, 
rather complicated in general and the problem may be quite difficult in 
practice. It 
would be interesting to investigate this further.

\section{The fully split case}

\subsection{Twist by $O(p)$}

	Let $(E,p)$ be an elliptic curve with a marked point  
and $V$ a holomorphic bundle of degree zero and rank $r$ on $E$.
Let $V':=V\otimes O(p)$.

	We say that $V$ is {\em fully split} if there exists a decomposition 
$V=\oplus_{j=1..r}{L_j}$ of $V$ into a direct sum of line bundles $L_j$.

We present an algorithm for determining whether a given
degree zero holomorphic vector bundle $V$ is semistable {\em and} fully split.
The algorithm requires explicit knowledge of $H^0(V')$ and
allows for the \det ermination of the line bundles $L_j$ up to 
holomorphic equivalence.

\begin{Theorem}	
\label{theorem_fully_split}

	Let $V$ be a degree zero holomorphic vector bundle of rank $r$ 
over $E$. Let $R_e:=R_e(V'),~r_e:=\dim_{\C}R_e,~K_e:=K_e(V')$ and 
$d_e:=\dim_{\C}K_e$ for any $e \in E$.

The following statements are equivalent :

(a) V is semistable and fully split 

(b) $V'$ satisfies all of the following conditions :

	(b0) $h^0(V')=r$

	(b1) The set $S:={\cal Z}(V')= \{ t \in E | r_t < r \}$ is finite 
	
	(b2) For all $t \in S$, all holomorphic sections of $V'$ belonging to 
	$K_t - {0}$ have degree $1$ at $t$

	(b3) For each $t \in S $ we have $V'_t = R_t \oplus N_t$

	(b4) We have $H^0(V')=\oplus_{t \in S}{K_t}$

(c) $V'$ satisfies (b0),(b1), (b4) and the condition that there exits a 
basis $(s_1..s_r)$ of $H^0(V')$ such that :

(b23) $\deg s_1\wedge s_2 \wedge ... \wedge s_r (t)= d_t,\forall t \in S$

	Moreover, in this case we have 
	$V \approx \oplus_{t \in S}{O(t-p)^{\oplus d_t}}$.

\end{Theorem}

Note that if $(b23)$ holds for a basis of $H^0(V')$ then it will hold 
for any other basis.

{\em Proof:}

{\em Show that  (a) implies (b):}
 
Assume $(a)$ holds and write $V=\oplus_{i=1..r}{L_i}$ with 
$L_i \in \Pic^0(E)$.
Then $L_i\approx O(q_i -p)$ $(q_i \in  E)$ and $V'=\oplus_{i=1..r}{L'_i}$, 
with $L'_i=L_i \otimes O(p)\approx O(q_i)$.

We know that $(b0)$ holds by Lemma \ref{cohom_dim}.
Let $s_i \in H^0(L'_i)-\{0\}$. Then $s_1 ... s_r$ is a $\C$-basis of $H^0(V')$.
We obviously have $S=\cup_{i=1..r}{\{q_i\}}$, so $(b1)$ holds.
Since $V'$ is semistable of slope $1$ we see that $(b2)$ also holds (cf. 
the remark before Proposition \ref{isom}).

Now suppose there are two distinct points $t_1,t_2 \in S$ such that 
$K_{t_1}\cap K_{t_2} \neq \{0\}$. Let $s \in K_{t_1}\cap K_{t_2}$.
Then $s(t_1)=s(t_2)=0$ and, since $s$ is regular, we must have $\deg L_s\geq2$,
which contradicts semistability of $V'$. Thus the sum $\sum_{t \in S}{K_t}$
is direct. On the other hand, any $s \in H^0(V')$ is a linear combination
$s =\sum_{i=1..r}{\alpha_i s_i}$ $(\alpha_i \in \C)$. Since $s_i \in K_{q_i}$,
we have $ s \in \sum_{t \in S}{K_t}$. Thus $(b4)$ holds.

To show $(b3)$, note that $L'_i=L_{s_i}$ (in the notation of subsection 2.1). 
Fixing $t\in S$, we clearly have 
$R_t=\oplus_{i; q_i\neq t}{(L'_i)_t}$, $K_t=\oplus_{i; q_i = t}{H^0(L'_i)}$ 
and $N_t=\oplus_{i; q_i=t}{(L'_i)_t}$.
Since $V'_t=\oplus_{i=1..r}{(L'_i)_t}$, we have $V'_t=R_t \oplus N_t$.

{\em Show that (b) implies (a):}

Let $d_t:=\dim_{\C}K_t$ $(t \in S)$.
By $(b4)$, we can choose a $\C$-basis $(s^{(t)}_j)_{t \in S, j=1..d_t}$ 
of $H^0(V')$ such that $(s^{(t)}_{j})_{j=1..d_t}$ is a $\C$ -basis of 
$K_t$ for each $t \in S$. By $(b4)$ and $(b2)$, each section $s^{(t)}_i$
has exactly one zero on $E$, namely at $t$, and this zero is simple   
(the unicity of this zero easily follows from $(b4)$). Therefore the line
bundles $L^{(t)}_j:=L_{s^{(t)}_j}$ have degree $1$ and we have
$L^{(t)}_j \approx O(t)$. In particular, for all $j=1..d_t$ we have 
$s^{(t)}_j(t)=0$ and ${\hat s}^{(t)}_j (t) \neq 0, \forall j=1..d_t $ 
and $s^{(t)}_j(t')\neq 0, \forall t' \in S - \{t \}$.
Moreover, $(b3)$ implies that $s^{(t')}_j(t) (t' \in S -\{t\},j=1..d_{t'})$
and ${\hat s}^{(t)}_j(t) (j=1..d_t)$ form a basis of $V'_t$. Therefore, we
have $V'_t = \oplus_{t' \in S, j=1..d_{t'}}{(L^{(t')}_j)_t}$, 
$\forall t \in S$.
On the other hand, for all $e \in E - S$ we have $\dim_{\C}R_e=r$. 
Since  $(s^{(t)}_j(e))_{t \in S, j=1..d_t}$ obviously generate $R_e$ 
and since $(b4)$ implies that 
$\Card\{s^{(t)}_j | t \in S, j=1..d_t\}=r$, it must be the case that 
$(s^{t}_j(e))_{t \in S, j=1..d_t}$ is a $\C$-basis of $V'_e$, 
for all $e \in S -t$. Therefore, we also have 
$V'_e = \oplus_{t \in S, i=1..d_t}{(L^{(t)}_i)_e}$, for $e \in E-S$.
Therefore, 
$V'=\oplus_{t \in S, j=1..d_t}{L^{(t)}_j}$. Since each component of this sum
has slope $1$, it follows that $V'$ is semistable and of slope $1$,
while $V=V' \otimes O(-p)$ is semistable and of degree zero. 
We also have $V'\approx \oplus_{t \in S}{O(t)^{d_t}}$ and 
$V\approx \oplus_{t \in S}{O(t-p)^{d_t}}$.

{\em Show that (b) and (c) are equivalent}

For this, assume that $(b0)$,
$(b1)$ and $(b4)$ hold. Then we show that $(b2)$ and $(b3)$ together are 
equivalent to $(b23)$. 

Remember that $\deg s_1\wedge...\wedge s_r(e)$ does not depend on the 
choice of the $\C$-basis of $H^0(V')$. Enumerating  
$S=\{t_1..t_k\}$ we can assume 
that $(s_i)_{d_1+...+d_{j-1}+1\le i \le d_1+...+d_j}$ is a $\C$-basis 
of $K_{t_j}$ for all $j=1..k$. Since the argument is similar for each 
$j$, let us focus on $t_1:=t$. Then $s_1...s_{d_t}$ is a 
basis of $K_t$ and for $i=1..d_t$ we have $s_i(e)=z\sigma_i(e)$ for all 
$e$ close to $t$, where $\sigma_i$ are local holomorphic sections 
of $V'$ around $t$.

{\em Claim 1}: If $(b0)$, $(b1)$ and $(b4)$ hold then 
$s_{d_t+1}(t)...s_r(t)$ is a basis of $R_t$.

Indeed, since $s_1(t)=...s_{d_t}(t)=0$, we clearly have that 
$s_{d_t+1}(t)...s_r(t)$ generate $R_t$. If $\alpha_{d_t+1}s_{d_t+1}(t)+
... + \alpha_rs_r(t)=0$ is zero a linear combination, then 
the section $s:= \alpha_{d_t+1}s_{d_t+1}+... + \alpha_r s_r$ of $V'$ 
vanishes at $t$ so that it belongs to $K_t$. Since our basis $s_1...s_r$ 
is `adapted' to the decomposition $(b4)$, $s$ then gives an element 
of $K_t \cap (\sum_{t' \in S -\{t\}}{K_{t'}})$, which must be zero 
since the sum in $(b4)$ is direct. Since $s_{d_t+1}...s_r$ are $\C$-linearly 
independent, this implies that $\alpha_{d_t+1}=..=\alpha_r=0$. Thus 
$s_{d_t+1}(t)...s_r(t)$ are linearly independent and the claim is proven.

{\em Claim 2}: If $(b0)$, $(b1)$ and $(b4)$ hold then the following are 
equivalent:

($\alpha$) $(b2)$ holds at $t$

($\beta$) $\sigma_1(t)...\sigma_{d_t}(t)$ are linearly independent

In this case, $\sigma_1(t)...\sigma_{d_t}(t)$ form a basis of $N_t$.

To prove this, first assume that $(b2)$ holds at $t$. Consider 
a zero linear combination 
$\alpha_1\sigma_1(t) +...+\alpha_{d_t}\sigma_{d_t}(t)=0$.
If the section $s:=\alpha_1s_1(t) +...+\alpha_{d_t}s_{d_t}$ would be nonzero, 
then it would have vanishing degree at least $2$ at $t$. This would contradict
$(b2)$. Therefore, we must have $s=0$ and $\alpha_1=..\alpha_{d_t}=0$. 
This proves that ($\alpha$) implies ($\beta$). 

Now assume that $(\beta)$ holds and consider a section $s \in K_t -\{ 0\}$. 
Then $s=\alpha_1s_1(t) +...+\alpha_{d_t}s_{d_t}$ for some $\alpha_i \in \C$ 
so that $s(e)=z(\alpha_1\sigma_1(e) +...+\alpha_{d_t}\sigma_{d_t}(e))$ 
for $e$ close to $t$. Since $s$ is not the zero section, at least one 
$\alpha_i$ is nonzero and $(\beta)$ implies that  
$\alpha_1\sigma_1(t) +...+\alpha_{d_t}\sigma_{d_t}(t)$ is nonzero. Thus 
$s$ has degree $1$ at $t$ and $(\alpha)$ holds. 

Assume that the equivalent conditions $(\alpha)$, $(\beta)$ hold and 
show that  $\sigma_1(t) ... \sigma_{d_t}(t)$ generate $N_t$.
We have $N_t:=<A>$, where $A:=\{{\hat s}(t)| s \in K_t\}$. 
If $s \in K_t-\{0\}$, the above arguments show that ${\hat s}(t)$ 
belongs to $<\sigma_1(t) ... \sigma_{d_t}(t)>$, and this is also 
trivially true for $s=0$ (since ${\hat s}(t)=0$ by definition in this case). 
Therefore we have $A \subset <\sigma_1(t) ... \sigma_{d_t}(t)>$ 
and $\sigma_1(t) ... \sigma_{d_t}(t)$ generate $N_t$. This finishes the proof 
of Claim 2. 

Now return to the proof of the theorem. 
Since $s_1(e) \wedge ... \wedge s_r(e)=
z(\sigma_1(e) \wedge ... \wedge \sigma_{d_t}(e) \wedge s_{d_t+1}(e)\wedge ... 
\wedge s_r(e))$ for $e$ close to $t$, $(b23)$ is equivelent to the statement 
that $\sigma_1(t) ...\sigma_{d_t}(t), s_{d_t+1}(t)... s_r(t)$ is a basis 
of $V'$. By Claim 1, linear independence of $s_{d_t+1}(t)... s_r(t)$ is 
automatic and $<s_{d_t+1}(t)... s_r(t)>=R_t$. By Claim 2, linear 
independence of $\sigma_1(t) ...\sigma_{d_t}(t)$ is equivalent to 
$(b2)$ and in this case $<\sigma_1(t) ...\sigma_{d_t}(t)>=N_t$. 
Then $<\sigma_1(t) ...\sigma_{d_t}(t), s_{d_t+1}(t)... s_r(t)>=V'_t$ 
is equivalent to $(b3)$. $\Box$

Let us explain how one can test $(b4)$. Suppose that $(b0),(b1)$ hold and let 
$s_1...s_r$ be an arbitrary  $\C$-basis of $H^0(V')$. 
For each $t \in S$, consider the $d_t$ -dimensional
subspace $P_t$ of $\C^r$ of linear relations among $s_1(t) ... s_r(t)$:

$P_t:=\{a:=(a_1 ... a_r) \in \C^r | a_1 s_1(t) + .. + a_r s_r(t)=0 \}$

Choose vectors $a^{(t,j)}\in \C^r$ 
$ (t \in S, j=1..d_t)$ such that, for each $t \in S$
$(a^{(t,j)})_{j=1..d_t}$ is a basis of $P_t$. 
Let $\zeta^{(t,j)}:=\sum_{i=1..r}{a^{(t,j)}_i~s_i} \in H^0(V')$. 
Then $(\zeta^{(t,j)})_{j=1..d_t}$ is a basis of $K_t$ for all $t \in S$.
In particular, we have $d_t=\dim_{\C}P_t$. 
Clearly  $(b4)$ is equivalent to the condition: 
\be 
\C^r=\oplus_{t \in S}{P_t}
\ee

Chosing an enumeration $S=\{t_i|i=1..k\}$ of $S$, 
we can form a matrix $A\in Mat(d,r,C)$, whose lines are given by the 
vectors $(a^{(t_i,j)})_{i=1..k, j=1..d_t}$. 
Then  $(b4)$ is equivalent to the conditions $d=r$ and $\det A \neq 0$.
Therefore, we obtain the following 

\

Algorithm :

\

Suppose $V$ is a rank $r$ and degree zero holomorphic vector bundle over $E$.
Let $p \in E$ arbitrary and define $V':=V\otimes O(p)$.

\

Step 1:

	Obtain a basis $(s_1 ... s_n)$ of $H^0(V')$.

\

Step 2:

	If $n \neq r $ then $V$ is not semistable. Otherwise, continue with 
Step 3.

\

Step 3:

	Let $\delta := s_1 \wedge ... \wedge s_r \in H^0(\Lambda^r V')$.
	If $\delta = 0 $ then $V$ is not semistable (this follows from 
        the main theorem in section 2). Otherwise, the set
	$S:=\supp (\delta)$
	is finite. In this case, enumerate $S=\{t_1 .. t_k\}$ and continue with
	Step 4.

\

Step 4 : 

	For each $t \in S$, determine $d_t=\dim_{\C}K_t$\footnote{
In general we can determine $d_t$ as $d_t=\dim_{\C}P_t$. In the monad case, 
$V'$ has a natural embedding into a direct sum of line bundles and 
$d_t$ can be determined directly by considering the rank of a matrix of 
sections as in \cite{CGL}}. 
Then $V'$ is semistable and fully split iff each of 
the following conditions is satisfied:

(a) $\sum_{t \in S}{d_t}=r$

(b) $\deg s_1\wedge ... \wedge s_r(t)=d_t$ for all $t \in S$

(c) The matrix $A$ is nonsingular

In this case, we have $V' \approx \oplus_{t \in S}{O(t)^{\oplus d_t}}$. 
In particular, the spectral divisor of $V$ is given by :
\be
\Sigma_{V}=\sum_{t \in S}{d_t~t}= (s_1 \wedge ... \wedge s_r)
\ee
Note that $\supp \Sigma_V =S$.

\subsection{More general twists}

Let $V$ be a fully split semistable vector bundle of degree zero over $E$.
Then $V=\oplus_{j=1..r}{L_j}$ with $L_j \in \Pic^0(E)$. 

Let $D=p_1+...+p_h$ be an effective divisor on $E$, where $p_1...p_h$ are 
{\em mutually distinct} points on $E$.
We use $p_1$ as a base point of $E$. Then we can write 
$L_j\approx O(q_j-p_1)$ with $q_j \in E$. 
Define:
\be
V':=V \otimes O(D)=\oplus_{j=1..r}{L'_j}
\ee
where $L'_j:=L_j \otimes O(D)\approx O(q_j+p_2+...+p_h)$.

Since $\deg L'_j=h$, we have $h^0(L'_j)=h$ and $h^0(V')=rh$.
The Riemann-Roch theorem gives $h^1(V')=0$. The spectral divisor 
of $V$ is $\Sigma_V=\sum_{j=1..r}{q_j}$. Let $S:=\supp \Sigma_V$. 
For each $q \in S$, let  $S_q=\{j \in \{1..r\} | q_j=q\}$
and $d_q:={\rm Card}S_q$.
Then $L'_j \approx O(q+p_2+..+p_h)$ for all $j \in S_q$.

\begin{Lemma}
\label{lemma_S}
Let $q \in E$ be arbitrary. The set :
\be
G_q(D):=\{s \in H^0(O(q+p_2+...+p_h)) | s(p_j)=0, \forall j=2..h\}
\ee
is a one-dimensional subspace of the $\C$-vector space 
$H^0(O(q+p_2+...+p_h))$.
Moreover, for any $s \in S_q(D)-\{0\}$ we have :
\be
(s)=q+p_2+p_3+...+p_h
\ee
\end{Lemma}

{\em Proof:}
Obviously the zero section belongs to $G_q(D)$.
Now let $s \in G_q(D) -\{0\}$. Since $s(p_2)=...=s(p_h)=0$, we have :
\be
\label{foo2}
(s)=D_s +p_2+...+p_h
\ee
with $D_s$ an effective divisor. Since $s \in H^0(O(q+p_2+...+p_h))$, 
we have $\deg (s)=\deg (q+p_2+...+p_h)=h$. But $\deg (s)=\deg D_s +(h-1)$ by 
(\ref{foo2}). Thus $\deg D_s$=1. Since $D_s$ is effective this implies $D_s=q'$
for some $q' \in E$.
On the other hand, $s \in  H^0(O(q+p_2+...+p_h))$ implies 
$(s)\sim q+p_2+...+p_h$, where $\sim$ denotes linear equivalence. Together 
with (\ref{foo2}), this gives $q' \sim q$. If $q' \neq q$, this would imply 
$E \approx \P^1$ by a classical theorem. Thus we must have $q'=q$ and 
$(s)=q+p_2+...+p_h$ for all $(s) \in G_q(D)-\{0\}$. By a standard argument
this implies that any $s' \in G_q(D)-\{0\}$ is of the form $s'=\lambda s$ 
with $\lambda \in \C^*$ a constant. Thus $G_q(D)$ is a one dimensional 
$\C$-vector space.
$\Box$

Let :
\be
G_j:=\{s \in H^0(L'_j) | s(p_2)=...=s(p_h)=0 \}\approx G_{q_j}(D)
\ee
($j=1..r$). 
By Lemma \ref{lemma_S}, $G_j$ are one-dimensional subspaces of $H^0(V')$. 
Define :
\be
G:=\{s \in H^0(V') | s(p_2)=...=s(p_h)=0\} \subset H^0(V')
\ee
and:
\be
G(q)=\oplus_{j \in S_q}{G_j}\subset G
\ee
(for all $q \in S$).

\begin{Proposition}
We have :
\be
\label{grad}
G=\oplus_{j=1..r}{G_j}=\oplus_{q \in S}{G(q)}
\ee
In particular, $G$ is an $r$-dimensional subspace of the $rh$ dimensional 
$\C$-vector space $H^0(V')$.
Moreover, for any $s\in G-\{0\}$ we have the alternative :

Either

(a) $(s)=q+p_2+...+p_h$ for some $q\in S$

or

(b)  $(s)=p_2+...+p_h$

\noindent If $(a)$ holds then $s \in G(q)$ for some $q \in S$, 
while if $(b)$ holds then $s \in G-\cup_{q \in S}{G(q)}$.
\end{Proposition}

Here the equalities are between divisors and {\em not} between divisor 
classes. That is, the equality in $(a)$ and $(b)$ is to be taken at face value
and {\em not} in the sense of linear equivalence.

{\em Proof:}

Since $V'=\oplus_{j=1..r}{L'_j}$, the statement $G=\oplus_{j=1..r}{G_j}$ is 
obvious.

Now let $s \in G-\{0\}$. By Proposition \ref{degree_bound}, we have: 
\be
\label{deg}
\deg (s)\le \mu(V')=h
\ee
Since $(s)$ is effective and $p_2...p_r \in \supp (s)$,  
there are only two possibilities :

(a) $(s)=q+p_2+...+p_h$ for some $q \in E$ (note that $q$ may belong to the 
set $\{ p_2...p_h\}$), and in this case $\deg (s)=h$

(b) $(s)=p_2+...+p_h$, and in this case $\deg (s)=h-1$.

\noindent Using $G=\oplus_{j=1..r}G_j$, we can write:
\be
\label{dir_sum}
s=\oplus_{j=1..r}{s_j}
\ee 
where $s_j \in G_j$. 
From $s_j \in G_j \subset H^0(L'_j)$, we obtain 
$(s_j)=q_j+p_2+...+p_h ~{\rm \ unless \ } s_j=0$.
Since (\ref{dir_sum}) is a direct sum, we have :
\be
\label{ineq}
(s_j) \geq (s) {\rm \ unless \ } s_j=0
\ee
for all $j=1..r$ \footnote{This means that $\deg (s_j)(e) \ge \deg (s)(e)$ for 
all $e \in E$.}. Indeed, $s$ can have a zero of order $m$ at $e\in E$ 
iff each $s_j$ has a zero of order at least $m$ at $e$. 

In case $(a)$, (\ref{ineq}) shows that $(s_j)=(s)$ for all $j=1..r$ with 
$s_j$ different from zero. This set is nonvoid iff $q \in S$ and in this 
case we obtain $s=\sum_{j \in S_q}{s_j} \in G(q)$. 

In case $(b)$, we cannot have $s \in G(q)$ for any $q$, since obviously this 
would imply $(s) \ge q+p_2+...+p_h$, a contradiction. $\Box$

\begin{Definition}
A $\C$-basis $\sigma_1...\sigma_r$ of $G$ is called {\em canonical} if 
$\deg \sigma_j=h$, for each $j=1..r$.
\end{Definition}

By the previous proposition, 
a basis of $G$ is canonical iff it is adapted to the 
graduation (\ref{grad}) of $G$, i.e. iff it is of the form 
$(\sigma^{q}_j)_{q \in S, j=1..d_q}$ with $(\sigma^{q}_j)_{j=1..d_q}$ 
bases of $G(q)$.

\begin{Corrolary}
Let $s_1...s_r$ be an arbitrary basis of $G$. Then 
the spectral divisor of $V$ is given by :
\be
\Sigma_V=(s_1\wedge ....\wedge s_r)-r(p_2+...+p_d)
\ee
\end{Corrolary}
\Pf Indeed, if $\sigma_1...\sigma_r$ is a canonical basis of $G$ then we have 
$(s_1\wedge ... \wedge s_r)=(\sigma_1\wedge ... \wedge \sigma_r)= 
\sum_{j=1..r}{q_j}+r(p_2+...+p_r)=\Sigma_V+r(p_2+...+p_r)$. $\Box$

This reduces the problem of determining the spectral divisor of 
$V$ to finding a basis of $G$. In the monad case, 
that can be easily accomplished 
by an obvious modification of the methods of \cite{CGL}.

It is now straightforward to formulate an analogue of Theorem 
\ref{theorem_fully_split}, in which $H^0(V')$ is replaced by $G$, 
whose proof is almost identical. Since this brings no new concepts to bear, 
we will not insist. 

There is a also a relatively straightforward generalization of the above 
to the non-fully-split case. A detailed statement would be rather 
lengthy and will not be given here.

\bigbreak\bigskip\bigskip\centerline{{\bf Aknowledgements}}\nobreak

The author would like to thank T.~M.~Chiang for  valuable comments on the 
manuscript. This work was supported by the DOE grant DE-FG02-92ER40699B  
and by a C.U. Fister Fellowship.

\end{document}